\documentclass[11pt,a4paper]{article}
\usepackage{jheppub}
\usepackage{amsmath}
\usepackage[most]{tcolorbox}
\usepackage{dsfont}
\usepackage{ulem}
\usepackage{natbib}
\usepackage{xcolor}
\usepackage[hang,flushmargin]{footmisc}
\usepackage{tikz-cd}
\usepackage{enumitem}
\makeatletter\renewcommand{\@biblabel}[1]{#1.}\makeatother

\newtcolorbox{empheqboxed}{colback=gray!20, 
 colframe=white,
 width=\textwidth,
 sharpish corners,
 top=0mm, 
 bottom=0pt
}

\def\be{\begin{equation}}
\def\ee{\end{equation}}
\def\i{{\rm i}}
\def\ti{{\rm i}}
\def\e{{\rm e}}
\def\d{{\rm d}}
\def\U{{\rm U}}
\def\q{{\mathfrak{q}}}
\def\p{{\mathfrak{p}}}
\def\t{{\mathfrak{t}}}

\def\S{{\mathbb{S}}}
\def\C{{\mathbb{C}}}

\newcommand{\nn}{\nonumber}

\newcommand{\mb}[1]{\mathbf{#1}}

\def\ket#1{{ |#1\rangle}}
\def\bra#1{{ \langle#1|}}
\def\braket #1#2{{ \langle #1|#2 \rangle}}

\def\res#1{\mathop \text{Res}\limits_{#1}}

\preprint{
{\small{\textsf{UUITP-37/18}}}}

\title{3d Mirror Symmetry from S-duality}

\author{Fabrizio Nieri$^a$, Yiwen Pan$^b$ and Maxim Zabzine$^a$}
\affiliation{$^a$Department of Physics and Astronomy, Uppsala University,\\
Box 516, SE-75120 Uppsala, Sweden. \\
\\
$^b$ School of Physics, Sun Yat-Sen University, Guangzhou, Guangdong, China
}

\emailAdd{fb.nieri@gmail.com}
\emailAdd{panyw5@mail.sysu.edu.cn}
\emailAdd{maxim.zabzine@physics.uu.se}

\abstract{We consider type IIB $\textrm{SL}(2,\mathbb{Z})$ symmetry to relate the partition functions of different 5d supersymmetric Abelian linear quiver Yang-Mills theories in the $\Omega$-background and squashed $\mathbb{S}^5$ background. By Higgsing S-dual theories, we extract new and old 3d mirror pairs. Generically, the Higgsing procedure yields 3d defects on intersecting spaces, and we derive new hyperbolic integral identities expressing the equivalence of the squashed $\S^3$ partition functions with additional degrees of freedom on the $\S^1$ intersection.   
}


\begin{document}
\maketitle
\flushbottom

\section{Introduction}

One of the most beautiful features in the family of 3d gauge theories with $\mathcal{N}=4$ supersymmetry is the existence of mirror symmetry \cite{Intriligator:1996ex}. When 3d supersymmetric gauge theories admit brane constructions through D3 branes suspended between $(p,q)$ branes \cite{Hanany:1996ie,Kitao:1998mf,Aharony:1997bh,Aharony:1997ju,Kol:1998cf}, mirror symmetry can be understood from the $\textrm{SL}(2,\mathbb{Z})$ symmetry of type IIB string theory. From the QFT perspective, mirror symmetry is deeply related to S-duality of the boundary conditions in 4d $\mathcal{N}=4$ supersymmetric Yang-Mills theory (SYM) \cite{Gaiotto:2008ak}, and for Abelian theories it can also be traced back to the existence of a natural $\textrm{SL}(2,\mathbb{Z})$ action on path integrals (functional Fourier transform) \cite{Witten:2003ya,Kapustin:1999ha}. For non-Abelian theories, this action can be implemented at the level of localized partition functions \cite{Gulotta:2011si,Assel:2014awa}. Moreover, the class of 3d $\mathcal{N}=4$ theories can be deformed in many interesting ways to $\mathcal{N}=2$, such as the inclusion of masses, Fayet-Iliopoulos (FI) parameters for Abelian factors in the gauge group, or superpotential terms. While the reduced supersymmetry implies a weaker control over the dynamics, mirror-like dualities are known to exist for a long time \cite{Aharony:1997gp,Aharony:1997bx,deBoer:1997kr}. Lately, this has been a very active research field, and significant progress is made possible thanks to the careful analysis of (monopole) superpotentials \cite{Aharony:2013dha,Aharony:2013kma,Amariti:2014iza,Benvenuti:2016wet,Benini:2017dud,Benvenuti:2017kud,Benvenuti:2017lle,Giacomelli:2017vgk}. In many cases, the IR equivalence of proposed dual pairs has been tested using the exact evaluation of supersymmetric observables through localization, such as the (squashed) $\mathbb{S}^3$ partition function \cite{Kapustin:2009kz,Hama:2011ea}. In fact, over the past few years, the results of supersymmetric localization (see e.g. \cite{Pestun:2016zxk} for a review) have been systematically exploited to predict and test dual pairs. 

In this paper, we continue the study of 3d dualities inherited from the $\textrm{SL}(2,\mathbb{Z})$ symmetry of type IIB string theory. Our strategy is to consider first 5d $\mathcal{N}=1$ SYM theories with unitary gauge groups engineered by $(p,q)$-webs in type IIB string theory in which the $\textrm{SL}(2,\mathbb{Z})$ action can be manifestly realized, for instance, through the exchange of D5 and NS5 branes (a.k.a. the fiber-base or S-duality \cite{Katz:1997eq,Bao:2011rc}). Secondly, we engineer codimension 2 defects of the parent 5d theories by the Higgsing procedure \cite{Gaiotto:2012xa,Gaiotto:2014ina}, and in simple configurations we can identify candidate 3d mirror pairs (this is the perspective also adopted in \cite{Zenkevich:2017ylb,Zenkevich:2018xx1,Zenkevich:2018xx2}). In order to be able to explicitly test their IR equivalence through the exact evaluation and comparison of the partition functions, we focus on 5d Abelian linear quivers in which the instanton corrections can be easily resummed \cite{Nekrasov:2008kza}. In fact, the fiber-base dual picture of such theories provides a very simple duality frame for the resulting 3d theories, which look free. Our reference example is 5d SQED with one fundamental and one anti-fundamental flavors and its fiber-base dual. From this very simple example, we can already extract non-trivial dualities for 3d non-Abelian theories. One of our main results is indeed a {\it non-Abelian version of the basic SQED/XYZ duality}. Remarkably, this duality  has implicitly appeared in \cite{Aghaei:2017xqe} (at the level of the squashed $\S^3$ partition function) as an intermediate step to test the mirror dual of $(A_1,A_{2n-1})$ Argyres-Douglas (AD) theories reduced to 3d, which has been shown to follow from an involved cascade of sequential confinement and mirror symmetry \cite{Benvenuti:2017lle,Benvenuti:2017kud} starting from the 3d reduction of the 4d ``Lagrangian" description \cite{Maruyoshi:2016tqk,Maruyoshi:2016aim}. Here, we provide a first principle derivation of this crucial bridge from the 5d physics viewpoint.

Another motivation for this paper comes from the recent studies of supersymmetric gauge theories on {\it intersecting spaces} \cite{Pan:2015hza,Gomis:2016ljm,Pan:2016fbl,Nekrasov:2016qym,Nekrasov:2016ydq,Nieri:2017ntx,Nieri:2018ghd}. In our case, we are interested in pairs of 3d theories supported on two codimension 2 orthogonal spaces in the ambient 5d space (which we take to be either the $\Omega$-background $\C^2_{\q,\t^{-1}}\times \mathbb{S}^1$ or the squashed $\mathbb{S}^5$ \cite{Lossev:1997bz,Moore:1998et,Moore:1997dj,Losev:1997tp,Nekrasov:2002qd,Nekrasov:2003rj,Kim:2012qf,Hosomichi:2012ek,Imamura:2012bm,Lockhart:2012vp,Kallen:2012va}), interacting along a common codimension 4 locus ($\mathbb{S}^1$) where additional degrees of freedom live. A natural question is whether 3d mirror symmetry survives in these more complicated configurations, and we can successfully generalize and test some of the old and the newly proposed dualities in this more refined setup too by studying the relevant compact and non-compact space partition functions.

The rest of the paper is organized as follows. In section \ref{sec:5dinst}, we review instanton partition functions of 5d Abelian linear quiver theories on $\C^2_{\q,\t^{-1}}\times \mathbb{S}^1$ through the refined topological vertex, exploiting their $(p,q)$-web realization in type IIB string theory or M-theory on toric Calabi-Yau 3-folds. In particular, the slicing invariance of the refined topological vertex implies the equivalence of supersymmetric partition functions of different looking field theories (duality frames) associated to the same string geometry. Two of the duality frames are exactly related by S-duality in type IIB, but we also discuss another one. In section \ref{sec:3dmirror}, we extract candidate 3d mirror pairs by following the Higgsings of the parent 5d theories across different duality frames, and compare the resulting partition functions. For special Higgsings, the 3d theories live on a single component codimension 2 subspace in the 5d ambient space, in which case we reproduce known results and propose a new mirror pair which is a non-Abelian version of the basic SQED/XYZ. However, we show that generic Higgsings produce 3d/1d coupled theories which live on distinct codimension 2 subspaces mutually intersecting along codimension 4 loci, and we generalize and test the dualities in these cases too. In section \ref{sec:discuss}, we discuss further our results and outline possible applications and extensions for future research. In appendix \ref{app:special}, we collect the definitions of the special functions which we use throughout the paper. In appendix \ref{app:computations} and \ref{app:S3partitionfunctions}, we present few technical definitions and derivations. In appendix \ref{sec:topvertex}, we collect useful information and notation of the refined topological vertex.

\section{5d instanton partition functions}\label{sec:5dinst}

In this section, we review the instanton partition functions  of 5d Abelian linear quiver theories with unitary gauge groups in the $\Omega$-background, usually denoted by $\C^2_{\q,\t^{-1}}\times \mathbb{S}^1$. The geometric engineering of these theories through $(p,q)$-webs in type IIB string theory or M-theory on toric Calabi-Yau 3-folds \cite{Aharony:1997bh,Aharony:1997ju,Kol:1998cf,Intriligator:1997pq,Leung:1997tw,Hollowood:2003cv} allows us to perform the various computations using the topological vertex formalism \cite{Aganagic:2003db,Iqbal:2004ne,Awata:2005fa,Iqbal:2007ii}. In this paper, we mainly follow the conventions of \cite{Awata:2008ed}, summarized in appendix \ref{sec:topvertex}. In a nutshell, in any toric diagram there is a frame in which one associates internal white arrows which point in the same (preferred/instanton) direction and correspond to unitary gauge groups, with the ranks determined by their number in each segment (one in this paper); consecutive gauge groups are coupled through bi-fundamental hypers, while non-compact white arrows correspond to (anti-)fundamental  hypers. 

\begin{figure}[t]
  \centering
  \includegraphics[width=0.25\textwidth]{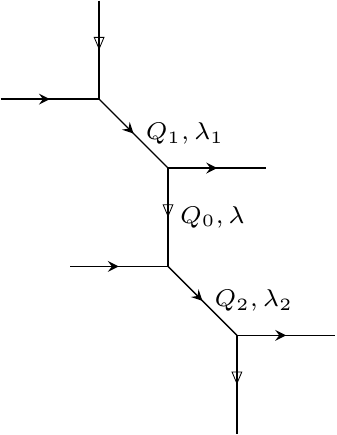}
  \includegraphics[width=0.25\textwidth]{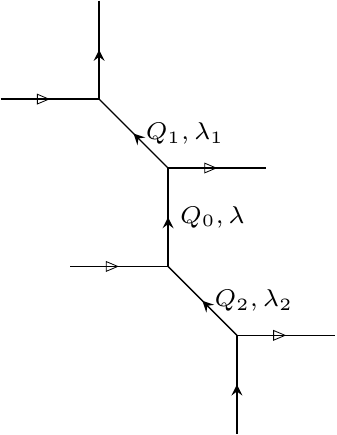}
  \includegraphics[width=0.25\textwidth]{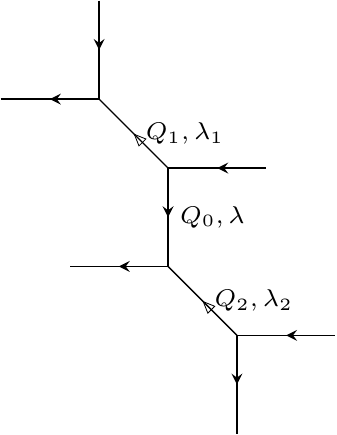}
  \caption{From left to right, the diagrams correspond to 5d ${\rm U}(1)$ gauge theories with 2 hypers, a free theory, and ${\rm U}(1)^2$ quiver theory with one bi-fundamental hyper multiplet. \label{three-diagrams}}
\end{figure}

Our reference examples are the diagrams listed in Figure \ref{three-diagrams}. By explicit computation, it is easy to verify that the associated topological amplitudes correspond respectively to the instanton partition functions of: $i)$ the $\textrm{U}(1)$ theory with one fundamental and one anti-fundamental hypers (SQED); $ii)$ the theory of four free hypers and ``resummed instantons", which will be simply referred to as the ``free theory"; $iii$) the $\textrm{U}(1)\times \textrm{U}(1)$ theory with one bi-fundamental hyper. This agrees with the thumb rule mentioned above.

The first diagram, corresponding to the $\textrm{U}(1)$ theory, has amplitude
\begin{align}\label{eq:Z1}
  Z_1 \equiv & \ \Bigg[\prod_{i=1}^2 \frac{1}{(Q_i \p^{1/2}; \q, \t^{-1})_\infty} \Bigg]
  \sum_\lambda (\p^{-1/2}Q_0)^{|\lambda|} \frac{N_{\emptyset \lambda} (Q_1 \p^{1/2}; \q,\t^{-1})  N_{\lambda \emptyset} (Q_2 \p^{1/2}; \q,\t^{-1})}{N_{\lambda\lambda}(1; \q,\t^{-1})} ~,
\end{align}
where $\p\equiv \q \t^{-1}$. The prefactor in front of the instanton sum can be identified with the perturbative or 1-loop contribution. We refer to appendix \ref{app:special} for the definition of $q$-Pochhammer symbols and Nekrasov's function.

The second diagram, corresponding to the free theory, has amplitude given by
\begin{align}
  Z_2 &\equiv  \frac{Z_{\rm Resummed}(Q_0,Q_1,Q_2)}{\Big[\prod_{i=1}^2(Q_i \p^{1/2}; \q,\t^{-1})_\infty (Q_0 \p^{1/2}; \q, \t^{-1})_\infty (Q_0 Q_1 Q_2 \p^{1/2} ; \q, \t^{-1})_\infty\Big]}~.\label{Z2}
\end{align}
Notice that the term in brackets contains the same perturbative contribution as before, and the whole bracket represents the contribution of four free hypers. However, the resummation of instantons has also produced the factor 
\be
Z_{\rm Resummed}(Q_0,Q_1,Q_2)\equiv (Q_0 Q_1 ;\q, \t^{-1})_\infty(Q_0 Q_2 \p;\q, \t^{-1})_\infty~,
\ee
which, being in the numerator, looks like the contribution of some exotic matter. Here, we simply take it as a computational result.\footnote{These contributions, sometimes called non-full spin content, are better explained in 6d \cite{Bao:2013pwa,Taki:2014pba}.}

Finally, the third diagram, corresponding to the $\textrm{U}(1) \times \textrm{U}(1)$ theory, has amplitude
\begin{align}
  Z_3 \equiv \Bigg[\frac{1}{(Q_0 \p^{1/2};\q, \t^{-1})_\infty}\Bigg] \sum_{\lambda_1, \lambda_2} (\p^{-1/2}Q_1)^{|\lambda_1|} (\p^{-1/2}Q_2)^{|\lambda_2|}\frac{N_{\lambda_1 \lambda_2} (Q_0\p^{1/2};\q,\t^{-1})}{N_{\lambda_1\lambda_1}(1;\q,\t^{-1}) N_{\lambda_2\lambda_2}(1;\q,\t^{-1})}\ .
\end{align}
The prefactor in front of the instanton sum can be identified with the perturbative contribution of the bi-fundamental hyper.

The above computation can be generalized to more complicated toric diagrams. For instance, a strip of $2N$ vertices can be associated to three QFT frames, corresponding respectively to: $i$) the $\textrm{U}(1)^{N - 1}$ theory coupled to $N-2$ bi-fundamentals, one fundamental at first node and one anti-fundamental at last node; $ii$) the theory of $2N$ free hypers and ``resummed instantons"; $iii$) the $\textrm{U}(1)^{N}$ theory coupled to $N-1$ bi-fundamentals. A similar {\it triality} relation among distinct gauge theories has been recently obtained also in 6d \cite{Bastian:2017ary,Bastian:2018dfu}.

\subsection{Duality frames}

\begin{figure}[h]
  \centering
  \includegraphics[width=0.8\textwidth]{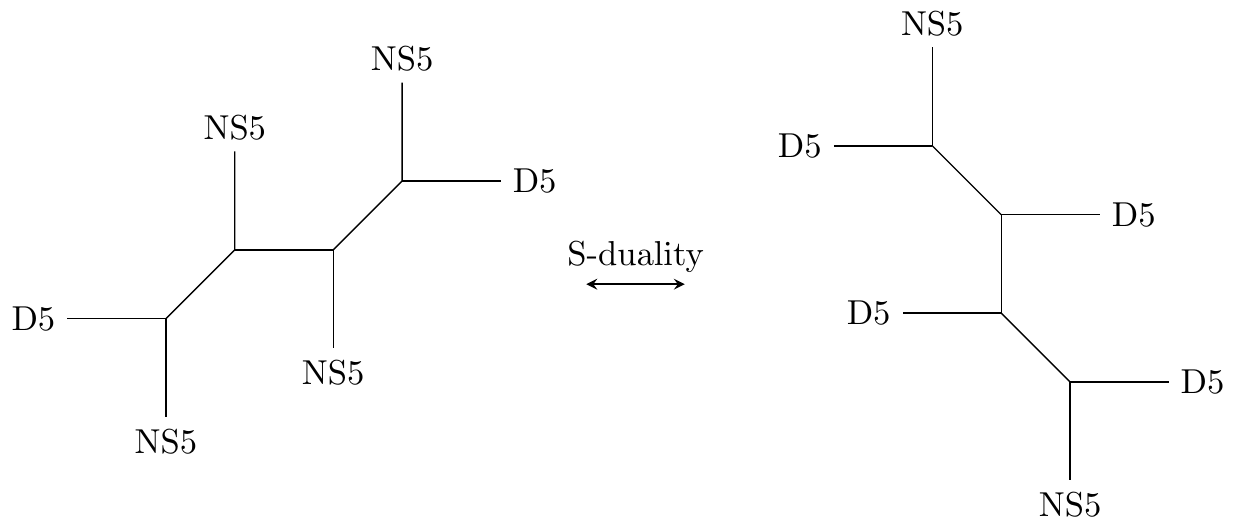}
  \caption{S-duality between the two $(p,q)$-webs. \label{S-duality}}
\end{figure}

The three configurations in Figure \ref{three-diagrams} share the same toric diagram. In fact, they all give equivalent amplitudes. Let us start by focusing on the first two diagrams in Figure \ref{three-diagrams}. They can be understood as two different $(p,q)$-webs related by S-duality in type IIB string theory, under which D5 and NS5 branes are exchanged. Upon a clockwise rotation by 90 degrees, the S-duality is represented by Figure \ref{S-duality}.
Since D5s correspond to horizontal $(1,0)$ branes, NS5s correspond to vertical $(0,1)$ branes and diagonal segments correspond to $(1,1)$ branes, the duality map is indeed represented by the $S$ element in $\mathrm{SL}(2,\mathbb{Z})$ acting on the $(p,q)$ charge vectors. In this particularly simple example, we can explicitly check the invariance of the amplitude. We can expand $Z_1$ and $Z_2$ in series of $Q_0$, and both $Z_1$ and $Z_2$ equal
\begin{multline}
  Z_1  =  \ \Bigg[    \prod_{i=1}^2 \frac{1}{(Q_i \p^{1/2}; \q,\t^{-1})}    \Bigg] \Bigg[   1 + \frac{Q_0}{(\q-1)(\t-1)} (\q Q_2 + \t Q_1 - (\q\t)^{1/2}(1+Q_1 Q_2)) + \ldots   \Bigg]=\\
  =  \ \Bigg[    \prod_{i=1}^2 \frac{1}{(Q_i \p^{1/2}; \q,\t^{-1})}    \Bigg]\Bigg[ 1  - \frac{(\q^{1/2}Q_2 - \t^{1/2})(\t^{1/2}Q_1 - \q^{1/2})}{(1 - \q)(1 - \t)} Q_0+ \ldots  \Bigg]=Z_2\ ,
\end{multline}
confirming one of the predictions. More generally, $(p,q)$-webs constructed by gluing vertically $N$ copies of the left diagram in Figure \ref{S-duality} or constructed by gluing horizontally $N$ copies of the right diagram are S-dual to each other and hence give equivalent amplitudes: the former is nothing but the $\textrm{U}(N)$ SQCD, while the latter is a linear $\textrm{U}(2)^{N-1}$ quiver with bi-fundamental hypers between gauge nodes and one fundamental and one anti-fundamental hypers at each ends.

The third duality frame is related to the first one by a clockwise rotation by 45 degree of the $(1,1)$ branes, which acts as the $STS^{-1}$ element in $\mathrm{SL}(2,\mathbb{Z})$ on the $(p,q)$ charge vector. One can verify that $Z_1=Z_2=Z_3$ using the identities of \cite{Awata:2008ed}.

The map between the K{\"a}hler parameters of the string geometry and the physical masses and coupling constants of the gauge theory depends on the duality frame. First of all, it is convenient to introduce exponential variables
\be\label{sec:mapsIN}
\q\equiv \e^{2\pi\i\beta\epsilon_1}~,\qquad \t\equiv \e^{-2\pi\i\beta\epsilon_2}~,\qquad Q_{1,2} \equiv \e^{2\pi\i\beta a_{1,2}}~,\qquad Q_0\equiv \e^{2\pi\i\beta a_0}~,
\ee
where $\beta$ measures the $\S^1$ radius. In the frame corresponding to the $\textrm{U}(1)$ theory, we can identify  
\be
a_1 \equiv -\i(\Sigma - \tilde M) \ , \qquad  a_2  \equiv \i(\Sigma - M) ~,\qquad 
Q_0(Q_1Q_2)^{1/2}\equiv \e^{-2\pi\beta\frac{8\pi^2}{g^2}}\ ,
\ee
where $M,\tilde M$ are the 5d fundamental and anti-fundamental masses, $\Sigma$ is the v.e.v. of the vector multiplet scalar and $g$ is the YM coupling.

Similarly, on the ${\rm U}(1)\times {\rm U}(1)$ side we can identify
\be\label{sec:mapsOUT}
a_0\equiv \i(\Sigma_{12}-M_{\rm bif})~,\quad
\p^{-1/2}Q_{1,2}\equiv \e^{-2\pi\beta\frac{8\pi^2}{g^2_{1,2}}}~,
\ee
where $M_{\rm bif}$ is the 5d bi-fundamental mass, $\Sigma_{12}\equiv\Sigma_1-\Sigma_2$ and $\Sigma_{1,2}$ are the v.e.v.'s of the vector multiplet scalars and $g_{1,2}$ are the YM couplings.

\subsection{$\mathbb{S}^5$ partition functions}
In this section, we use the refined topological string/Nekrasov partition functions in the various duality frames to write $\mathbb{S}^5$ partition functions related by type IIB $\textrm{SL}(2,\mathbb{Z})$ transformations. The study of compact space partition functions is useful because one can get rid of subtleties related to boundary conditions, at the price of introducing an integration over some modulus. The round $\mathbb{S}^5 \equiv \{(z_1, z_2, z_3) \in \mathbb{C}^3| \ |z_1|^2 +  |z_2|^2 +  |z_3|^2 = 1\}$ admits a toric $\U(1)^3$ action given by $z_\alpha\to \e^{\i\varphi_\alpha}z_\alpha$. Denoting by $e_\alpha$ the corresponding vector fields, the vector ${\footnotesize \textsc{R}}=e_1+e_2+e_3$ is the so-called Reeb vector, and it describes the Hopf fibration $\U(1)\rightarrow\S^5\rightarrow\mathbb{CP}^2$. A useful generalization is obtained by replacing the Reeb vector with ${\footnotesize \textsc{R}}=\omega_1e_1+\omega_2e_2+\omega_3e_3$ ($\omega_i\in\mathbb{R}_{>0}$), and the resulting manifold is referred to as the squashed $\S^5$ and the $\omega$'s as squashing (or equivariant) parameters. We refer to \cite{Qiu:2014oqa} for further details of this geometry.
 
The partition functions of 5d $\mathcal{N} = 1$ gauge theories on the (squashed) $\S^5$ can be computed via localization. In the Coulomb branch localization scheme \cite{Kim:2012qf,Hosomichi:2012ek,Imamura:2012bm,Lockhart:2012vp} (as opposed to the Higgs branch scheme \cite{Pan:2014bwa,Pan:2015hza,Nieri:2018ghd}), the result is given in terms of a matrix-like integral over the constant vector multiplet scalar in the Cartan subalgebra of the gauge group. It is known that the integrand can be constructed by gluing three Nekrasov partition functions \cite{Kim:2012qf,Lockhart:2012vp,Nieri:2013vba,Qiu:2013aga,Qiu:2014oqa}, one for each fixed point of the toric action on $\mathbb{CP}^2$, with equivariant parameters $\epsilon_{1,2}$ and radius $\beta$ of the $\Omega$-background related to (complexified) squashing parameters. For each of the fixed points labeled by $\alpha=1,2,3$, where the space looks like a copy of $\mathbb{C}^2_{\q,\t^{-1}}\times\mathbb{S}^1_\beta$, we can choose
\begin{align}\label{eq:tableequiv}
  \begin{tabular}{c | c c c}
    & $1$ & $2$ & $3$\\
    \hline
    $\epsilon_1$ &  $\omega_1 + \omega_2 $ &  $\omega_2 + \omega_3$ & $\omega_1 + \omega_3$  \\
    $\epsilon_2$  & $\omega_3$ & $\omega_1$& $\omega_2$\\
    $\beta$  & $1/\omega_1$ & $1/\omega_2$& $1/\omega_3$\\
  \end{tabular}~~~~~~~.
\end{align}

On the ${\rm U(1)}$ theory side (frame 1), the product of Nekrasov partition functions yields
\be
 \bigg| Z_1 \bigg|^3 \equiv \frac{\e^{-\frac{\i\pi}{6}\big(B_{33}(-\i (\Sigma - \tilde M) + \frac{\omega}{2})+B_{33}(\i (\Sigma -  M) + \frac{\omega}{2})\big)}}{S_3(-\i (\Sigma - \tilde M) + \frac{\omega}{2})S_3(\i (\Sigma - M) + \frac{\omega}{2})}  \ \bigg|  Z_{\text{inst} | {\rm U}(1)}^{\C^2 \times \S^1} (g; \Sigma, M, \tilde M) \bigg|^3 \ ,
\ee
where $\omega\equiv \omega_1+\omega_2+\omega_3$ and $|~\cdot~|^3$ denotes the product of three objects with parameters related by table (\ref{eq:tableequiv}). Notice that the 1-loop contributions have fused into triple Sine functions (and exponential factors) by using the definition (\ref{eq:facS3}).

On the free theory side (frame 2), the product of $q$-Pochhammer symbols yields 
\begin{multline}
\bigg|Z_2\bigg|^3  \equiv  \frac{\prod_\pm\e^{-\frac{\i\pi}{6}\big(B_{33}(\pm\frac{8 \pi ^2 \i}{g^2} + \frac{\i}{2}  (\tilde M- M) + \frac{\omega}{2})-B_{33}(\pm \frac{8 \pi ^2 \i}{g^2}  + \frac{\i}{2}  (M+\tilde M)- \i \Sigma)\big)}}{\e^{\frac{\i\pi}{6}\big(B_{33}(-\i (\Sigma - \tilde M) + \frac{\omega}{2})+B_{33}(\i (\Sigma - M) + \frac{\omega}{2})\big)}}\times\\
\times\frac{1}{S_3(-\i (\Sigma - \tilde M) + \frac{\omega}{2})S_3(\i (\Sigma - M) + \frac{\omega}{2})}\prod_\pm \frac{S_3(\pm \frac{8 \pi ^2 \i}{g^2} + \frac{\i}{2}  (M+\tilde M)-\i \Sigma )}{S_3(\pm\frac{8 \pi ^2 \i}{g^2} + \frac{\i}{2}  (\tilde M- M) + \frac{\omega}{2})} ~.
\end{multline}

Using $Z_1=Z_2$ (type IIB S-duality), after removing common exponential factors on both sides and integrating with the classical action, we can obtain the identity\footnote{The identity is actually stronger because it is really an identity even before taking the integral.}
\begin{align}
  Z^{\mathbb{S}^5}_\text{SQED}= Z^{\mathbb{S}^5}_\text{free}\ ,
  \label{SQED-free-duality}
\end{align}
where we defined the squashed $\mathbb{S}^5$ partition function of the  SQED by
\begin{align}\label{S5SQED}
  Z^{\mathbb{S}^5}_\text{SQED} &\equiv  \ \int \d\Sigma \, Z^{\rm cl}_\text{SQED}(g;\Sigma)\, Z^\text{1-loop}_\text{SQED}(\Sigma,M,\tilde M) \, \bigg| Z_{\text{inst} | {\rm U}(1)}^{\C^2 \times \S^1} (g; \Sigma, M, \tilde M) \bigg|^3~,
\end{align}
\begin{align}
 Z^{\rm cl}_\text{SQED}(g;\Sigma)&\equiv \e^{- \frac{8 \pi^3 }{\omega_1 \omega_2 \omega_3} \frac{\Sigma^2}{g^2}}~,\nn\\
 Z^\text{1-loop}_\text{SQED}(\Sigma,M,\tilde M)&\equiv  \frac{1}{S_3(-\i (\Sigma - \tilde M) + \frac{\omega}{2})S_3(\i (\Sigma - M) + \frac{\omega}{2})} \ , 
\end{align}  
and the ``Fourier-like transform" of the squashed $\S^5$ partition function of free theory  by
\begin{align}
  Z^{\mathbb{S}^5}_\text{free}& \equiv   \int \d\Sigma \,  \e^{\frac{8\pi ^3\i\Sigma(\i M + \i \tilde M -\omega)}{g^2 \omega_1 \omega_2 \omega_3}}\, Z_\text{Resummed}(\Sigma,M,\tilde M,g)\, Z^\text{1-loop}_\text{free}(\Sigma,M,\tilde M,g)~,
\end{align}
\begin{align}
Z_\text{Resummed}(\Sigma,M,\tilde M,g)&\equiv \e^{-\frac{8\pi ^3 (\i M - \frac{\omega}{2})(\i \tilde M - \frac{\omega}{2}) }{g^2 \omega_1 \omega_2 \omega_3}}\,\prod_\pm S_3(- \i \Sigma \pm \frac{8 \pi ^2 \i}{g^2} + \frac{\i}{2}  (M+\tilde M) )~,\nn\\
Z^\text{1-loop}_\text{free}(\Sigma,M,\tilde M,g)&\equiv \frac{1}{S_3(-\i (\Sigma - \tilde M) + \frac{\omega}{2})S_3(\i (\Sigma - M) + \frac{\omega}{2})}\times\nn\\
&~~~~~~~~~~~~~~~~~~~~~\times\prod_\pm \frac{1}{S_3(\pm\frac{8 \pi ^2 \i}{g^2} + \frac{\i}{2}  (\tilde M- M) + \frac{\omega}{2}) }~.
\end{align}
As for the $\Omega$-background case, we simply take this result as a computational fact and we do not attempt to give here a gauge theory interpretation, which is not needed for the purposes of this paper.

On the ${\rm U}(1)\times {\rm U}(1)$ side (frame 3), we can write
\be
\bigg| Z_3 \bigg|^3\equiv\frac{\e^{-\frac{\i\pi}{6}B_{33}(\i\Sigma_{12}-\i M_{\rm bif}+\frac{\omega}{2})}}{S_3(\i(\Sigma_{12}-M_{\rm bif})+\frac{\omega}{2})}\bigg| Z_{\text{inst} |{\rm U}(1)^2}^{\C^2 \times \S^1} (g_1,g_2; \Sigma_{12}, M_{\rm bif})\bigg|^3~,
\ee
and in order to reproduce the squashed $\mathbb{S}^5$ partition function we need to bring the exponential factor on the other side and integrate with  the classical action, namely 
\be
Z^{\mathbb{S}^5}_{\U(1)^2}\equiv\int\d\Sigma_1\d\Sigma_2\, Z^{\rm cl}_{\U(1)^2}(g_1,g_2;\Sigma_1,\Sigma_2)\, Z^\text{1-loop}_{\U(1)^2}(\Sigma_{12},M_{\rm bif})\bigg| Z_{\text{inst} |{\rm U}(1)^2}^{\C^2 \times \S^1} (g_1,g_2; \Sigma_{12}, M_{\rm bif})\bigg|^3~,
\ee
\begin{align}
Z^{\rm cl}_{\U(1)^2}(g_1,g_2;\Sigma_1,\Sigma_2)&\equiv \e^{-\frac{8\pi^3}{\omega_1\omega_2\omega_3}\left(\frac{\Sigma_1^2}{g_1^2}+\frac{\Sigma_2^2}{g_2^2}\right)}~,\nn\\
Z^\text{1-loop}_{\U(1)^2}(\Sigma_{12},M_{\rm bif})&\equiv \frac{1}{S_3(\i(\Sigma_{12}-M_{\rm bif})+\frac{\omega}{2})}~.
\end{align}

Substituting $Z_3=Z_2=Z_1$ and using the dictionary (\ref{sec:mapsIN})--(\ref{sec:mapsOUT}), one can obtain two more identities. In the following, we are going to focus on first one, namely type IIB S-duality in relation to 3d mirror symmetry.

\section{Mirror symmetry}\label{sec:3dmirror}

In this section, we will follow type IIB S-duality acting on 5d gauge theories, and extract mirror dual partition functions of 3d gauge theories defined on the squashed $\mathbb{S}^3$ or on the intersecting space $\mathbb{S}^3_{(1)} \cup \mathbb{S}^3_{(2)} \subset \mathbb{S}^5$.  The spheres $\mathbb{S}^3_{(\alpha)}$ are submanifolds associated to the equations $z_\alpha = 0$, $\alpha = 1,2,3$. We will focus on $\mathbb{S}^3_{(1)}$ and $\mathbb{S}^3_{(2)}$, which clearly  intersect transversally\footnote{The intersection is transversal from the perspective of the two $\mathbb{C}$'s in the two individual tubular neighborhoods $\mathbb{C} \times \mathbb{S}^1 \subset \S^3_{(1) \mathrm{ or } (2)}$. Put differently, the two complex planes intersect only at the origin.} along the circle $|z_3| = 1$. We will denote the squashing parameters of $\mathbb{S}^3_{(1)}$ and $\S^3_{(2)}$ by $b_{(1)} = \sqrt{\omega_2/\omega_3}$ and $b_{(2)} = \sqrt{\omega_1/\omega_3}$ respectively, and we will set $Q_{(\alpha)} \equiv b_{(\alpha)} + b_{(\alpha)}^{-1}$ as usual. We will review few aspects of gauge theories on this type of geometries in the following, while for further details we refer to \cite{Gomis:2016ljm,Pan:2016fbl}.

\subsection{Higgsing, residues and mirror symmetry}

\begin{figure}
  \centering
  \includegraphics[width=\textwidth]{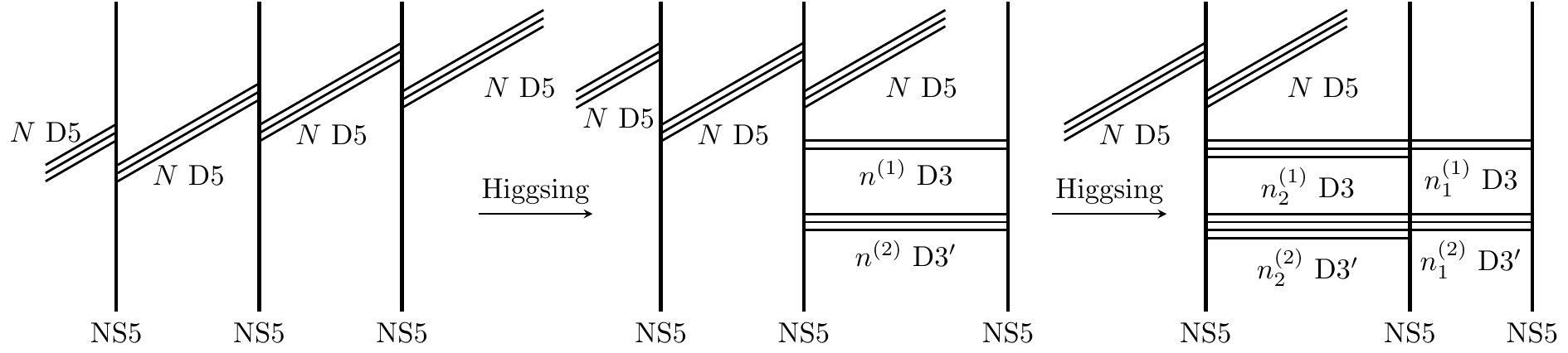}
  \caption{The brane moves of a simple type of Higgsing applied to a 5d linear quiver gauge theory. The NS5s fill the $012348$ directions, while the D5s fill the $012347$ directions, hence $78$ represents the $(p,q)$-plane (here we are slightly simplifying the picture). The D3s are stretched along the 6 direction and fill also the $012$ and/or $034$ directions, hence they all share a common direction and are supported on two orthogonal planes inside the 5-brane worldvolumes. \label{Higgsing-brane-construction}}
\end{figure}

Higgsing \cite{Gaiotto:2012xa,Gaiotto:2014ina} a higher dimensional bulk theory is an effective procedure for accessing lower dimensional supersymmetric theories that preserve half (or fewer) the supercharges that the bulk theory enjoys. More precisely, the resulting lower dimensional supersymmetric field theories are worldvolume theories of codimension 2 BPS defects inserted into the bulk theory. The procedure can be more easily described when there is a (flat space) brane construction. If the 5d theory $\mathcal{T}$ admits a construction in terms of an array of D5s suspended between parallel NS5s, for example when $\mathcal{T}$ is a unitary linear quiver gauge theory, then one type of Higgsing amounts to aligning the outermost flavor D5 with the adjacent gauge D5, and subsequently pulling the in-between NS5 away from the array while stretching a number of D3s. See Figure \ref{Higgsing-brane-construction} for an example .
 At the level of the compact space partition function, Higgsing $\mathcal{T}$ implies taking the residues at certain poles of the partition function as a meromorphic function of mass parameters. In practice, when the compact space partition function is written as a Coulomb branch integral, this is often equivalent to computing the residues of the integrand at a collection of poles of the perturbative determinant as a function of the v.e.v.'s of the scalars in the vector multiplet(s).

Let us consider the partition function of the SQED on the squashed $\mathbb{S}^5$  expressed as an integral as in (\ref{S5SQED}). In the following we will focus on the poles of $Z_\text{SQED}^\text{1-loop}(\Sigma)$ of the form
\be
\i (\Sigma - M)+\omega/2 = -{n^{(1)}}  \omega_1 - {n^{(2)}}\omega_2 - n^{(3)} \omega_3 ~,\quad n^{(\alpha)}\in\mathbb{Z}_{\geq 0}~.\label{poles}
\ee
It is sufficient to study the cases with $n^{(3)} = 0$ as they already demonstrate many core features of more general cases. The cases $n^{(3)} \neq 0$ are a straightforward generalization. As was extensively discussed in \cite{Pan:2016fbl}, the residue of the integrand can be organized into the partition function of a 5d/3d/1d coupled system. Indeed, upon taking the residue, a few things happen which we now summarize (we refer to \cite{Pan:2016fbl} for a full account, and to appendix \ref{app:computations} for the sketch of a slightly different derivation). The non-perturbative factors and the classical factor are simply evaluated at the pole $\Sigma^* = M +\i \omega/2+\i n^{(1)}\omega_1 + \i n^{(2)}\omega_2$. Because of the different $\S^1$ periodicities at the fixed points $|z_{\alpha}|=1$ labeled by $\alpha=1,2,3$ as in table (\ref{eq:tableequiv}), the $n^{(1)}$ dependence in the first instanton partition function drops out, and it only depends on $n^{(2)}$, and similarly the second depends only on $n^{(1)}$, while the third depends on both. Therefore, among the three instanton partition functions associated to the three fixed points, two simply reduce to the vortex partition functions of two SQCDAs with gauge groups ${\rm U}(n^{(2)})$ and ${\rm U}(n^{(1)})$ supported on $\big(\mathbb{C}_{t^{-1}} \times \mathbb{S}^1\big)_{(1)}$ and $\big(\mathbb{C}_\q \times \mathbb{S}^1\big)_{(2)}$ respectively, while the remaining one encodes the vortex partition functions of the two SQCDA,\footnote{We refer to the $\U(n)$ SYM theory coupled to $n_\text{f}$ fundamental, $n_\text{f}$ anti-fundamental and 1 adjoint chiral multiplets with SQCDA.} now supported on $\big(\mathbb{C}_\q \times \mathbb{S}^1\big)_{(3)}$ and $\big(\mathbb{C}_{t^{-1}} \times \mathbb{S}^1\big)_{(3)}$ respectively, and their intricate interaction along the common $\S^1$ at the origin. Schematically, we have the reduction
\begin{multline}
   \ Z_\text{SQED}^{\rm cl}(\Sigma)\Big|Z_{\text{inst}|\text{SQED}}^{\mathbb{C}^2 \times \mathbb{S}^1}(\Sigma) \Big|^3  \xrightarrow{\Sigma = \Sigma^*}  \  Z^\text{cl}_{\rm SQED}(\Sigma^*)\times\\
   \times \underbrace{\Big(Z^{\C_{\t^{-1}}\times\S^1}_{{\rm vortex}|\U(n^{(2)})}\Big)_{(1)}}_{\Big(Z_{\text{inst} | {\rm U}(1)}^{\C^2 \times \S^1}(\Sigma^*)\Big)_{(1)} }\underbrace{\Big(Z^{\C_{\q}\times\S^1}_{{\rm vortex}|\U(n^{(2)})}Z^{\S^1\ (+\text{``extra"})}_{{\rm int.|\U(n^{(2)})\times\U(n^{(1)})}}Z^{\C_{\t^{-1}}\times\S^1}_{{\rm vortex}|\U(n^{(1)})}\Big)_{(3)}}_{\Big(Z_{\text{inst} | {\rm U}(1)}^{\C^2 \times \S^1}(\Sigma^*)\Big)_{(3)} }\underbrace{\Big(Z^{\C_{\q}\times\S^1}_{{\rm vortex}|\U(n^{(1)})}\Big)_{(2)}}_{\Big(Z_{\text{inst} | {\rm U}(1)}^{\C^2 \times \S^1}(\Sigma^*)\Big)_{(2)} }~,
\end{multline}
where the ``extra" factors are remnants that will eventually cancel out in the final result. Also, the residue of the 1-loop factors can be simplified to 
\begin{align}\label{eq:1loopRes}
  \res{\Sigma \to \Sigma^*} Z^\text{1-loop}_\text{SQED}(\Sigma) = Z_\text{HM}^{\mathbb{S}^5}(\tilde M - M) \times \ldots ~,
\end{align}
where $Z_\text{HM}^{\mathbb{S}^5}(M) \equiv S_3(\i M + \omega/2)^{-1}$ denotes the 1-loop determinant of a free hyper of mass $M$ on the squashed $\mathbb{S}^5$, while the dots denote 1-loop determinant factors similar to those which would arise in a Higgs branch localization computation of SQCDAs on each $\S^3$ \cite{Benini:2013yva,Gomis:2014eya}, plus interaction terms. Because of the form of the $\q$, $\t$ parameters at each fixed point and the 3d holomorphic block factorization of $\S^3$ partition functions \cite{Pasquetti:2011fj,Beem:2012mb,Taki:2013opa}, one can readily understand that the above reduction describes the partition function of the combined system of two SQCDA on $\S^3_{(1)}$ and $\S^3_{(2)}$, interacting through additional degrees of freedom at the common $\S^1$.\footnote{Notice that a generalization to the three-component subspace $\S^3_{(1)}\cup \S^3_{(2)}\cup \S^3_{(3)}\subset \S^5$ features in the Higgs branch localization on $\S^5$ \cite{Nieri:2018ghd}.}

To make our life easier when dealing with the defect theories, it is convenient to recast the above Higgs branch-like representation of the partition function sketched above, into a Coulomb branch-like integral, making the structure of the worldvolume theories manifest. This is possible thanks to the following non-trivial observation: one can reorganize all the (intricated) factors into an elegant matrix integral, namely
\begin{empheqboxed}

\textbf{Proposition 1 (residues).}
\begin{multline}\label{matrixmodel}
\frac{\res{\Sigma \to \Sigma^*}Z_\text{SQED}^{\rm cl}(\Sigma)Z^\text{1-loop}_\text{SQED}(\Sigma)\Big|Z_{\text{inst}|\text{SQED}}^{\mathbb{C}^2 \times \mathbb{S}^1}(\Sigma) \Big|^3}{Z^{\rm cl}_\text{SQED}(M+\i\omega/2) \, Z_\text{HM}^{\mathbb{S}^5}(\tilde M-M)}= \\
=\int \prod_{\alpha=1}^2\prod_{a=1}^{n_a^{(\alpha)}}\frac{\d\sigma^{(\alpha)}_a}{2\pi\i \, n^{(\alpha)}!}\, Z^{\mathbb{S}^3_{(1)}}_{{\rm U}(n^{(1)})\text{-SQCDA}}(\sigma^{(1)}) Z^{\mathbb{S}^1}_\text{1d chiral}(\sigma^{(1)},\sigma^{(2)})Z^{\mathbb{S}^3_{(2)}}_{{\rm U}(n^{(2)})\text{-SQCDA}}(\sigma^{(2)})\equiv \\
\equiv  Z^{\mathbb{S}^3_{(1)} \cup\, \mathbb{S}^3_{(2)}}_\text{${\rm U}(n^{(1)})$-SQCDA $\cup$ ${\rm U}(n^{(2)})$-SQCDA}~.
\end{multline}
\end{empheqboxed}
The explicit expression of the integrand of the matrix model on the r.h.s. can be found in appendix \ref{app:S3partitionfunctions}, and the definition of the integral is given by the Jeffrey-Kirwan prescription discussed in \cite{Pan:2016fbl}. The proof of this equality relies on formal manipulations of Nekrasov's functions and brute force computational checks, as briefly explained in appendix \ref{app:computations}.

\begin{figure}
  \centering
  \includegraphics[width=0.35\textwidth]{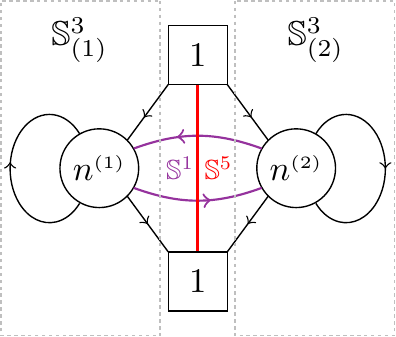}
  \caption{
    The quiver structure of the 5d/3d/1d theory describing the 5d theory in the presence of intersecting codimension 2 defects. The hyper or chiral multiplets supported on spheres of different dimensions are indicated by their colors.
    \label{intersecting-SQCDA}
  }
\end{figure}

To summarize, the result of the residue computation can be naturally interpreted as the partition function of a free hyper multiplet on the squashed $\mathbb{S}^5$ in the presence of two BPS codimension 2 defects supported respectively on $\mathbb{S}^3_{(1)}$ and $\mathbb{S}^3_{(2)}$ which intersect along a common $\mathbb{S}^1 = \mathbb{S}^3_{(1)} \cap \mathbb{S}^3_{(2)}$. Each defect is characterized by its worldvolume theory, being 3d $\mathcal{N} = 2$ ${\rm U}(n^{(\alpha)})$-SQCDA with $\alpha=1,2$ respectively. It is crucial to emphasize that the two defect worldvolume theories interact at an $\mathbb{S}^1$, which harbors a pair of additional 1d $\mathcal{N} = 2$ chiral multiplets transforming in the bi-fundamental representation of the two 3d gauge groups. Figure \ref{intersecting-SQCDA} summarizes the quiver structure of the 5d/3d/1d coupled system.
Each SQCDA on $\mathbb{S}^3_{(\alpha)}$ contains one fundamental, one anti-fundamental and one adjoint chiral multiplet, of masses $m^{(\alpha)}$, $\tilde m^{(\alpha)}$ and $m_{\rm adj}^{(\alpha)}$ respectively. The FI term is turned on with coefficient $\zeta_{(\alpha)}$. These parameters can be identified with the 5d hyper multiplet masses and gauge coupling according to the dictionary
\begin{align}
   m^{(\alpha)} &= \lambda_\alpha \Big(M + \frac{  \i  }{  2  }(\omega + \omega_\alpha) \Big)~, &\quad \tilde m^{(\alpha)} &=  \lambda_\alpha \Big(\tilde M  + \frac{  \i  }{  2  }(\omega - \omega_\alpha)  \Big)~, \nonumber\\
  m_\text{adj}^{(\alpha)} &= \i \omega_{\alpha}\lambda_\alpha ~, &\quad \zeta_{(\alpha)} &= \frac{8\pi^2 \lambda_\alpha}{g^2} \label{mass-identification}~ ,
\end{align}
where $\lambda_\alpha \equiv \sqrt{\omega_\alpha/\omega_1 \omega_2 \omega_3}$. The 3d chiral multiplets $q, \tilde q$ couple to the bulk hyper multiplet $q_{\rm bulk}$\footnote{The 5d hyper multiplet has two scalars $q_{\rm bulk}, \tilde q_{\rm bulk}$. It can be decomposed into two 3d $\mathcal{N} = 2$ chiral multiplets, into which $q_{\rm bulk}$ and $\tilde q_{\rm bulk}$ enter separately.} via cubic superpotentials $q_{(1)}\tilde q_{(1)} q_{\rm bulk}$ and $q_{(2)}\tilde q_{(2)} q_{\rm bulk}$, leading to the mass relations
\be
 b_{(1)} m^{(1)} - b_{(2)} m ^{(2)} = \frac{\i}{2}(b_{(2)}^2 - b_{(1)}^2)\ ,\qquad b_{(1)} \tilde m^{(1)} - b_{(2)} \tilde m ^{(2)} = - \frac{\i}{2}(b_{(2)}^2 - b_{(1)}^2) ~,\nn
\ee
\be
 b_{(1)} m^{(1)}_\text{adj} - b_{(2)} m^{(2)}_\text{adj} = \i (b_{(2)}^2 - b_{(1)}^2)\ .
\ee
In other words, the theories on $\mathbb{S}^3_{(1)}$ and $\mathbb{S}^3_{(1)}$ share the same ${\rm U}(1)$ flavor group. The FI parameters in the two theories are also related by $(b\zeta)_{(1)} = (b\zeta)_{(2)}$, indicating that the two theories also share the same ${\rm U}(1)$ topological symmetry.

Now we are ready to extract candidate 3d  mirror pairs. The two sides of the fiber/base duality between frame 1 and 2 share the same poles in the integrand. In fact, the integral equality {\it trivially} follows from the equality of the integrand, and therefore by taking the residue at the same pole $\Sigma \to \Sigma^*$ on both sides (and dropping the common factors), we extract a family of {\it non-trivial} integral identities labeled by non-negative integers $n^{(1)}$ and $n^{(2)}$, namely
\begin{empheqboxed}
\textbf{Proposition 2 (master identity).}
\begin{multline}\label{master-equality}
Z^{\mathbb{S}^3_{(1)} \cup\, \mathbb{S}^3_{(2)}}_\text{$\U(n^{(1)})$-SQCDA $\cup$ $\U(n^{(2)})$-SQCDA}= \\
=  \exp\Bigg[ \frac{8\pi^3}{g^2 \omega_1 \omega_2 \omega_3}(- \i(M + \tilde M) + \omega) (n^{(1)}\omega_1 + n^{(2)}\omega_2) - \frac{16 \pi^3}{g^2 \omega_3}n^{(1)}n^{(2)}\Bigg]\times\\
\!\!\!\times\prod_{\alpha=1}^2\prod_{k=0}^{n^{(\alpha)}-1}\frac{S_2\Big(\i (\tilde M- M) + \omega + k \omega_\alpha|\omega_3, \omega_\gamma\Big)}{S_2\Big( - (k+1)\omega_\alpha)|\omega_3, \omega_\gamma \Big)\prod_\pm S_2\Big(\pm \frac{8\pi^2 \i}{g^2} + \frac{\i}{2}(\tilde M - M) + \frac{\omega}{2} + k \omega_\alpha|\omega_3, \omega_\gamma \Big)}\times \\
\times \prod_{k=1}^{n^{(1)}}\prod_{\ell=1}^{n^{(2)}}
\frac{\prod_\pm S_1 \Big(\pm \frac{8\pi^2 \i}{g^2} + \frac{\i}{2}(\tilde M - M) + \frac{\omega}{2} + (k-1) \omega_1 + (\ell-1) \omega_2|\omega_3 \Big)}{S_1\Big(- k \omega_1 - \ell \omega_2 | \omega_3\Big)S_1\Big(\i (\tilde M -  M) + \omega + (k-1) \omega_1 + (\ell-1) \omega_2 | \omega_3\Big)}\, ,
\end{multline}
\end{empheqboxed}
where $\gamma=1,2$ when $\alpha=2,1$. We refer to appendix \ref{app:special} for the definitions of the double Sine and single Sine functions. Notice that this mathematical identity, which we will refer to as the {\it master identity}, is new and provides a huge generalization of the hyperbolic identity in Theorem 5.6.8 of \cite{Fokko:2007}. The proof relies on formal manipulations of Nekrasov's functions and brute force computational checks. We will shortly see that these integral identities, derived from type IIB S-duality, capture 3d $\mathcal{N} = 2$ mirror symmetry on intersecting $\S^3$'s.

\subsection{Warming up: SQED/XYZ duality}

We begin with a warm-up exercise to see that the well-known Abelian mirror symmetry between 3d $\mathcal{N} = 2$ SQED and the XYZ model arises from the integral identities discussed above. For this, we consider $n^{(1)} = 1$ and $n^{(2)} = 0$. Upon substituting in (\ref{mass-identification}), the master equality (\ref{master-equality}) implies
\be
\frac{Z^{\mathbb{S}^3_{(1)}}_\text{$\U(1)$-SQCDA}}{s_b \Big(\frac{\i Q}{2} + m_\text{adj}\Big)_{(1)}}
=\Bigg[  \e^{ - \pi \i \zeta(m+\tilde m)}
   s_b \Big(\frac{\i Q}{2} + m - \tilde m \Big)s_b \Big(  - \zeta - \frac{m - \tilde m}{2} \Big)s_b \Big( + \zeta - \frac{m - \tilde m}{2} \Big) \Bigg]_{(1)}~.
\ee
We refer to appendix \ref{app:special} for the definition of the double sine function. This integral equality is nothing but the mirror symmetry relation between 3d $\mathcal{N} = 2$ SQED and the XYZ model at the level of  $\mathbb{S}^3_{(1)}$ partition functions. As expected, the complexified masses of the three free chiral multiplets in the XYZ model, namely (suppressing the label $^{(1)}$) 
\be
  m_{X ,Y} \equiv \pm \zeta - \frac{m - \tilde m}{2} - \frac{\i Q}{2} ~, \qquad m_Z\equiv m - \tilde m~ ,
\ee
satisfy \begin{align}
  m_X + m_Y + m_Z = - \i Q ~ ,
\end{align}
signaling the presence of the superpotential $XYZ$. On the SQED side, the additional 1-loop factor signals the presence of a decoupled chiral multiplet $\beta_1$ interacting with the adjoint chiral $\Phi$ through the superpotential $\beta_1\Phi$.

\subsection{Generalization: intersecting SQED/XYZ duality}

Now we are ready to generalize the mirror symmetry relation between the SQED and XYZ models to intersecting spheres. Dropping from both sides the common 1-loop factors like $\prod_{\alpha = 1}^2 s_b(\i Q/2 + m_\text{adj})_{(\alpha)}$, the master equality (\ref{master-equality}) with $n^{(1)} = n^{(2)} = 1$ implies a more involved integral identity, namely
\begin{multline}
 Z^{\S^3_{(1)} \cup\, \S^3_{(2)}}_\text{SQED $\cup$ SQED} =\\
  = \frac{
       \sin \frac{\i\pi}{2}  \Big( b_{(1)} m_X^{(1)} + b_{(2)} m_X^{(2)} \Big)
       \sin \frac{\i\pi}{2}  \Big( b_{(1)} m_Y^{(1)} + b_{(2)} m_Y^{(2)} \Big)
    }{
       \sin \i\pi \Big(b_{(1)} m^{(1)}_\text{adj} + b_{(2)}m^{(2)}_\text{adj} \Big)
       \sin \frac{\i\pi}{2} \Big( b_{(1)} (m_Z + m_\text{adj})^{(1)} + b_{(2)} (m_Z + m_\text{adj})^{(2)} \Big)
    } \times\\
  \times \prod_{\alpha=1}^2\Bigg[  \e^{ - \pi \ti \zeta (m + \tilde m) }
  s_{b}\bigg(\frac{\ti Q}{2} + m_Z \bigg)
  s_{b} \bigg( \frac{\i Q}{2} + m_X \bigg)
  s_{b} \bigg( \frac{\i Q}{2} + m_Y\bigg)\Bigg]_{(\alpha)} \ ,
\end{multline}
where the masses in the XYZ models are defined as usual by (we suppress the label $^{(\alpha)}$)
\begin{align}
  m_{X ,Y} \equiv \pm \zeta - \frac{m - \tilde m}{2} - \frac{\i Q}{2} ~, \qquad m_Z\equiv m - \tilde m~ .
\end{align}

The l.h.s. of the above identity is the partition function of two SQED on $\S^3_{(1)}$ and $\S^3_{(2)}$, coupled through a pair of 1d bi-fundamental chiral multiplets along the common $\S^1$ intersection. The r.h.s. can be naturally interpreted as the partition function of two XYZ models on $\S^3_{(1)}$ and $\S^3_{(2)}$, coupled to a pair of 1d free Fermi multiplets and another pair of 1d chiral multiplets on $\S^1$. The fact that the masses of the 1d multiplets are combinations of those of the 3d multiplets indicates the presence of a certain 1d superpotential that involves both the 3d and 1d chiral multiplets. As a result, the 1d multiplets are charged under the 3d global symmetries, in particular, the Fermi multiplets are charged under the 3d topological $\U(1)$ symmetry.

\subsection{Generalization: non-Abelian SQCDA/XYZ duality}

We can now move to discuss more interesting examples, generalizing the previous Abelian examples to non-Abelian gauge groups. Let us start by considering $n^{(1)}>0$, $n^{(2)}=0$, in which case the master equality specializes to
\begin{multline}\label{nonintersecting-nonabeilian-duality}
  Z^{\mathbb{S}^3_{(1)} }_\text{$\U(n^{(1)})$-SQCDA}=\\
= \Bigg[
        \e^{- \pi \i n \zeta (m + \tilde m) }
        \prod_{\mu = 0}^{n-1}
          s_b\Big( \frac{\i Q}{2} + m_{\Phi\mu} \Big)
          s_b\Big( \frac{\i Q}{2} + m_{Z\mu} \Big)
          s_b\Big(\frac{\i Q}{2} + m_{X\mu}\Big)
          s_b\Big(\frac{\i Q}{2} + m_{Y\mu}\Big)
      \Bigg]_{(1)} ~,
\end{multline}
where we used the shorthand notations (suppressing again the label $^{(1)}$)
\begin{align}
m_{X \mu} &\equiv \zeta - \frac{m - \tilde m}{2} - \frac{\i Q}{2} - \mu m_\text{adj}~,& m_{Y \mu} &\equiv  - \zeta - \frac{m - \tilde m}{2} - \frac{\i Q}{2} - (n - 1 - \mu) m_\text{adj}~,\nn\\
m_{Z\mu}&\equiv m - \tilde m + \mu m_\text{adj}~,& m_{\Phi\mu}& \equiv (\mu + 1) m_\text{adj} ~.
\end{align}
For convenience, we can reorganize the following products
\begin{align}
  & \prod_{\mu = 0}^{n-1}s_b\Big( \frac{\i Q}{2} + m_{\Phi\mu}\Big)
  = \frac{1}{ \prod_{\mu = 1}^{n } s_b\Big( \i Q/2 - \mu m_\text{adj} - \i Q \Big)}~, \\
  & \prod_{\mu = 0}^{n-1}s_b\Big( \frac{\i Q}{2} + m_{Z\mu} \Big) = \frac{
    s_b\Big( \i Q/2 + m - \tilde m + (n-1) m_\text{adj} \Big)}{
    \prod_{\mu = 0}^{n-2}s_b\Big( \i Q/2 - m + \tilde m - \mu m_\text{adj} -\i Q\Big)
  } \ ,
\end{align}
and move the denominators to the l.h.s. of (\ref{nonintersecting-nonabeilian-duality}). Defining the leftover mass on the r.h.s. 
\be
m_Z\equiv m_{Z(n-1)}= m - \tilde m + (n - 1)m_\text{adj}~,
\ee
one easily finds the masses satisfy
\begin{align}
  m_{X \mu} + m_{Y\mu} + m_Z = - \i Q \ , \qquad \mu = 0, \ldots, n-1 \ ,
\end{align}
which is compatible with the superpotential $\sum_{\mu = 0}^{n - 1} X_\mu Y_\mu Z$. On the l.h.s., the additional 1-loop factors are compatible with free chiral multiplets $\beta_\mu$ and $\gamma_\mu$ interacting with the adjoint chiral $\Phi$ and the quarks $q,\tilde q$ through the superpotential $\sum_{\mu=0}^{n-2}\gamma_\mu \tilde q\Phi^\mu q+\sum_{\mu=1}^{n}\beta_\mu \Phi^\mu$. 

\begin{figure}
  \centering
  \includegraphics[width=0.5\textwidth]{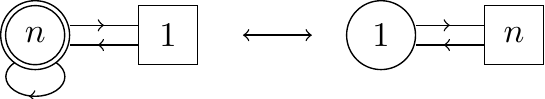}
  \caption{\label{SUn-duality}Another duality that can be obtained by integrating over the FI parameter, viewed as the ``mass'' for the topological $\U(1)$ symmetry.}
\end{figure}

The mathematical relation (\ref{nonintersecting-nonabeilian-duality}) has implicitly appeared in \cite{Aghaei:2017xqe} as an intermediate step to test another duality, involving the ${\rm SU}(n)$ theory coupled to one fundamental, one anti-fundamental and one adjoint chiral on the one hand, and the $\U(1)$ theory coupled to $n$ hypers on the other hand as shown in Figure \ref{SUn-duality}, which was motivated by the study of the mirror dual of $(A_1,A_{2n-1})$ AD theories reduced to 3d \cite{Benvenuti:2017lle,Benvenuti:2017kud}. This duality is simply related to ours by gauging the topological $\U(1)$. Hence, we have physically interpreted and derived both dualities as 3d $\mathcal{N}=2$ mirror symmetry descending from type IIB S-duality.

\subsection{Generalization: intersecting non-Abelian SQCDA/XYZ duality}

It is now straightforward to take the further generalization $n^{(1)},n^{(2)}>0$. In this case, the master identity yields
\begin{align}\label{intersecting-nonabeilian-duality}
  &Z^{\mathbb{S}^3_{(1)} \cup\, \mathbb{S}^3_{(2)}}_\text{$\U(n^{(1)})$-SQCDA $\cup$ $\U(n^{(2)})$-SQCDA}=\nn \\
&=\prod_{\alpha = 1}^2 \Bigg[
        \e^{- \pi \i n \zeta (m + \tilde m) }
        \prod_{\mu = 0}^{n-1}
          s_b\Big( \frac{\i Q}{2} + m_{\Phi\mu} \Big)
          s_b\Big( \frac{\i Q}{2} + m_{Z\mu} \Big)
          s_b\Big(\frac{\i Q}{2} + m_{X\mu}\Big)
          s_b\Big(\frac{\i Q}{2} + m_{Y\mu}\Big)
      \Bigg]_{(\alpha)} \times\nn\\
&    \times\!\!\!\! \prod_{\mu = 0}^{n^{(1)} - 1}\prod_{\nu = 0}^{n^{(2)} - 1} \!\frac{
       \sin \frac{\pi \i}{2} \Big(b_{(1)} (m_{X\mu} - m_{\Phi\mu})^{(1)} + b_{(2)}(m_{X\mu} - m_{\Phi\nu})^{(2)} + \i b_{(2)}^2 + \i b_{(1)}^2\Big)
      \Big(X \to Y\Big)
    }{
       \sin \pi \i \Big(
        b_{(1)}m_{\Phi\mu}^{(1)} + b_{(2)}m_{\Phi\nu}^{(2)}
      \Big)
       \sin \frac{\pi \i}{2} \Big(
        b_{(1)}(m_{Z\mu} + m_{\Phi\mu})^{(1)}
        + b_{(2)}(m_{Z\nu} + m_{\Phi\nu})^{(2)}
      \Big)
    }\ ,\nn\\
    &
\end{align}
where we used the same shorthand notations as before. We can reorganize the factors as we did in the previous subsection, and the difference compared to the previous result (besides the doubling of all factors) is the presence of the additional 1-loop contributions from the 1d matter living on the $\S^{(1)}$ intersection, represented by the last line. This picture provides the generalization of the non-Abelian SQCDA/XYZ duality to the more complicated geometry involving 1d degrees of freedom, and we have shown that it also descends from type IIB S-duality.

It is worth noting that one can further integrate over the FI parameters $\zeta_{(i)}$ to obtain the intersecting space version of the $\text{SU}(n)$-SQCDA/$\U(1)$ duality mentioned at the end of the last subsection. However, the fact that the FI parameters on each component space are related by $(b\zeta)_{(1)}=(b\zeta)_{(2)}$ implies integration with the constraint $\delta( \sum_{a=1}^{n^{(1)}}(b^{-1} \sigma_a)_{(1)} + \sum_{a=1}^{n^{(2)}}(b^{-1} \sigma_a)_{(2)})$, whose field theory interpretation remains unclear to us at the moment.

\subsection{Quiver gauge theories}

\begin{figure}
  \centering
  \includegraphics{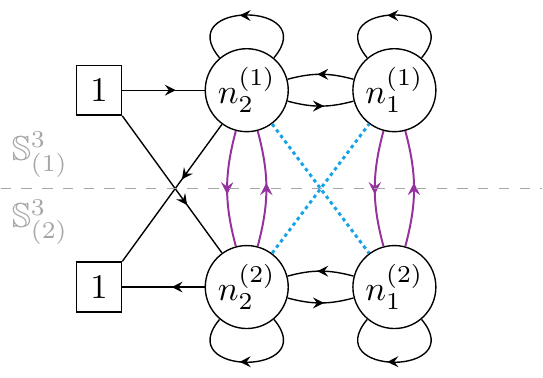}
  \caption{Quiver worldvolume theories of intersecting codimension 2 defects following from Higgsing twice. The purple arrows denote bi-fundamental 1d chiral multiplets, while the blue dotted lines denote 1d Fermi multiplets. \label{intersecting-quiver}}
\end{figure}

It is possible to generalize the above computations to quiver gauge theories. As shown in Figure \ref{Higgsing-brane-construction}, one could start from a 5d linear quiver gauge theory and engineer intersecting codimension 2 defects with quiver worldvolume theories by multiple Higgsings. For example, it is not hard to convince oneself that by Higgsing twice the 5d linear quiver gauge theory with two $\U(1)$ gauge nodes, one will obtain 3d quiver theories of the form depicted in Figure \ref{intersecting-quiver}.
It is possible to apply the Higgsing procedure by taking the residues of the resulting partition functions and their fiber/base dual, and repeat the computations in the previous discussions. However, the technical computations are more involved and we do not consider them here explicitly.

\section{Discussion and outlook}\label{sec:discuss}

In this paper, we have studied a class of 3d $\mathcal{N}=2$ non-Abelian gauge theories which can be realized as codimension 2 defects in the parent 5d $\mathcal{N}=1$ Abelian gauge theories, which in turn can be realized in type IIB string theory. Generically, the defect theories are not supported on a single component subspace, instead, they live on mutually orthogonal submanifolds intersecting at codimension 4 loci where additional degrees of freedom live. We have considered some implications of type IIB $\text{SL}(2,\mathbb{Z})$ symmetry for these systems, and we have generalized to this class of more complicated geometries the known fact that type IIB S-duality reduces to 3d mirror symmetry. Using the refined topological vertex, we have been able to test this idea in simple cases where the parent 5d gauge theory is simply the SQED with two flavors, while the dual 3d theories are SQCDA with two chirals and a generalized XYZ model. Interestingly enough, the QFT/string theory methods have also allowed us to physically explain existing integral identities in the math literature, and moreover, to derive new ones and interpret them as the equivalence of partition functions of mirror dual theories on (intersecting) squashed spheres.

Along the lines of this paper, one should also be able to study more complicated 5d theories and hence derive new or generalized 3d mirror pairs. As byproduct, one may also obtain new mathematical identities expressing the equivalence of dual partition functions. Moreover, what we have discussed in this paper is expected to have a higher dimensional lift \cite{Mironov:2016cyq} by considering 6d theories engineered by periodic $(p,q)$-webs \cite{Hollowood:2003cv,Hohenegger:2013ala,Haghighat:2013tka} and the resulting 4d/2d defect theories. 

Finally, it is worth noting that the type of 3d/1d defects that we have considered in this paper appear in the Higgs branch localization approach to SQCD on $\S^5$ \cite{Nieri:2018ghd}, whose partition functions are identified with correlators in the $q$-Virasoro modular triple \cite{Nieri:2017mod}. Therefore, another interesting route of investigation would be the study of type IIB $\text{SL}(2,\mathbb{Z})$ symmetry from the viewpoint of the BPS/CFT and 5d AGT correspondences \cite{Nekrasov:2012xe,Carlsson:2013jka,Nekrasov:2013xda,Nekrasov:2015wsu,Awata:2009ur,Awata:2010yy,Mironov:2011dk,Nieri:2013yra,Nedelin:2016gwu,Mitev:2014isa,Isachenkov:2014eya,Aganagic:2015cta,Kimura:2015rgi,Benvenuti:2016dcs,Aganagic:2017smx} and the DIM algebra \cite{Ding1997,miki:2007}, whose representation theory is known to govern the topological amplitudes associated to toric CY 3-folds or $(p,q)$-webs \cite{Mironov:2016yue,Awata:2016riz,Awata:2016mxc,Awata:2016bdm}. From this perspective, the $\text{SL}(2,\mathbb{Z})$ symmetry group is identified with the automorphism group of the DIM algebra, and it would be interesting to systematically study how different $q$-deformed correlators are related to each other. In turn, this perspective may give powerful tools for handling 3d mirror symmetry very efficiently. This is a topic which deserves further investigations, and in appendix \ref{sec:topvertex} we have collected few preliminary comments and background material  for the interested readers.

\acknowledgments
We thank S. Pasquetti for valuable comments and discussions. We also thank the Simons Center for Geometry and Physics (Stony Brook University) for hospitality during the program ``Localization Techniques in Quantum Field Theories", at which some of the research for this paper was performed. F.N. also thanks N. Haouzi and P. Koroteev for discussions, and UC Berkeley and UC Davis for hospitality. The research of F.N. and M.Z. is supported in part by Vetenskapsr\r{a}det under grant \#2014-5517, by the STINT grant and by the grant ``Geometry and Physics" from the Knut and Alice Wallenberg foundation. Y.P. is supported by the 100 Talents Program of Sun Yat-sen University under Grant No.74130-18831116.

\begin{appendix}
  \section{Special functions}\label{app:special}

In this appendix, we recall the definitions of several special functions which we use in the main body. Below, $r$ is a positive integer, and $\vec\omega\equiv(\omega_1, \ldots, \omega_r)$ is a collection of non-zero complex parameters. We frequently take $r = 1,2,3$ for concreteness. We refer to \cite{Narukawa:2003} for further details. 

The \textbf{multiple Bernoulli polynomials} $B_{r n}(X|\vec\omega)$ are defined by the generating function
\begin{align}
  \frac{t^r \e^{Xt}}{\prod_{i=1}^r \e^{\omega_i t} - 1} \equiv \sum_{m \geq 0} B_{rn}(X|\vec\omega) \frac{t^n}{n!}\;.
\end{align}
In particular, we use $B_{22}(X|\vec\omega)$ and $B_{33}(X|\vec\omega)$ in this note, and they are given explicitly by
\begin{align}
  B_{22}(X|\vec\omega) &\equiv  \ \frac{X^2}{\omega_1 \omega_2} - \frac{\omega_1 + \omega_2}{\omega_1 \omega_2} X + \frac{\omega_1^2 + \omega_2^2 + 3 \omega_1 \omega_2}{6\omega_1 \omega_2}\;,\\
  B_{33}(X|\vec\omega) &\equiv B_{33}(X)=\ \frac{X^3}{\omega_1 \omega_2 \omega_3} - \frac{3(\omega_1 + \omega_2 + \omega_3)}{2\omega_1 \omega_2 \omega_3} X^2  + \nn\\
  &+\frac{\omega_1^2 + \omega_2^2 + \omega_3^2 + 3 \omega_1 \omega_2 + 3 \omega_2 \omega_3 + 3 \omega_3 \omega_1}{2\omega_1 \omega_2 \omega_3}X+\nn\\
  & \qquad - \frac{(\omega_1 + \omega_2 + \omega_3)( \omega_1 \omega_2 +  \omega_2 \omega_3 + \omega_3 \omega_1)}{4\omega_1 \omega_2 \omega_3} \; .
\end{align}

The \textbf{$q$-Pochhammer symbols} are defined as
\begin{equation}
  (x;{q_1},\dots ,{q_r})_\infty \equiv \prod\limits_{{n_1},\ldots ,{n_r} = 0}^\infty  {(1 - x q_1^{{n_1}}\ldots q_r^{{n_r}})} \qquad {\text{when all }}|{q_i}| < 1\;.
\end{equation}
Other regions in the $q$-planes are defined through the replacements
\begin{equation}
  (x;{q_1},\dots ,{q_r})_\infty \to \frac{1}{{(q_i^{ - 1}x;{q_1},\ldots ,q_i^{ - 1},...,{q_r})_\infty}}\;.
\end{equation}

The \textbf{multiple Sine functions} $S_r(X|\vec\omega) $ can be defined by the $\zeta$-regularized product
\begin{align}
  S_r(X|\vec\omega) \simeq \prod_{m_1, \ldots, m_r \in \mathbb{N}} \Big(X + \sum_{i=1}^r m_i \omega_i \Big)^{(-1)^{r+1}}\Big( - X + \sum_{i=1}^r (m_i + 1) \omega_i \Big)\;.
\end{align}
$S_r(X|\vec\omega)$ is symmetric in all $\omega_i$, has the reflection property $S_r(X|\vec \omega) = S_r(\omega - X| \vec\omega)^{(-1)^{r+1}}$ for $\omega\equiv \omega_1 + \ldots + \omega_r$, the homogeneity property $S_r(\lambda X|\lambda \vec\omega)=S_r(X|\vec\omega)$ for $\lambda\in\mathbb{C}^\times$, and the shift property
\begin{align}
  S_r(X + \omega_i|\vec\omega) = \frac{S_r(X|\vec\omega)}{S_{r-1}(X|\widehat{ \omega})} \ , \quad \widehat{\omega} \equiv (\omega_1, \omega_{i-1},   \omega_{i+1}, \ldots, \omega_r)\ .\label{shift-property}
\end{align}
The \textbf{single Sine function} $S_1(X|\vec\omega)$ is simply defined as
\be
S_1(X|\vec\omega)\equiv 2\sin(\pi X/\omega_1)~.
\ee

The \textbf{double Sine function} $S_2(x|\vec\omega)$ enjoys a factorization property when $\operatorname{Im}(\omega_1/\omega_2) \ne 0$, namely
\begin{align}
  S_2(X |\vec\omega) = \e^{\frac{\ti \pi}{2} B_{22}(X|\vec\omega)} (\e^{2\pi \ti X/\omega_1}; \e^{2\pi \ti \omega_2/ \omega_1})_\infty(\e^{2\pi \ti X/\omega_2}; \e^{2\pi \ti \omega_1/ \omega_2})_\infty \ . \label{factorization-S2}
\end{align}
There is also a shifted version of the double Sine function which is often denoted by $s_b(X)$ where $b \equiv \sqrt{\omega_1/\omega_2}$, related to $S_2(X|\vec \omega)$ by
\begin{align}
  S_2(X|\vec\omega) \equiv s_b \bigg(  - \frac{\ti Q}{2} + \frac{\ti X}{\sqrt{\omega_1 \omega_2}} \bigg)~,
\end{align}
where $Q\equiv b+b^{-1}$. In terms of the double sine $s_b(x)$, the factorization is rewritten as
\begin{align}\label{factorization-sb}
  s_b\bigg(  -\frac{\ti Q}{2} + X  \bigg) &= \e^{\frac{\i\pi }{2} B_{22}(-\ti X| b, b^{-1})} (\e^{2\pi b X}; \e^{2\pi \ti b^2})_\infty(\e^{2\pi b^{-1} X}; \e^{2\pi \ti b^{-2}})_\infty \ .
\end{align}

The reflection property of $s_b(z)$ is simply
\begin{align}
  s_b (X)  s_b (-X) = 1\ .
\end{align}

The \textbf{triple Sine function } $S_3(X|\vec\omega)\equiv S_3(X)$ also has a useful factorization property. When $\operatorname{Im}(\omega_i/\omega_j) \ne 0$ for all $i \ne j$, then
\begin{align}\label{eq:facS3}
  S_3(X) = \e^{- \frac{\i\pi }{6} B_{33}(X)} \prod_{1\leq i\neq j\neq k\leq 3}\big( \e^{\frac{2\pi\i}{\omega_k}X}; \e^{2\pi\i\frac{\omega_i}{\omega_k}},\e^{2\pi\i\frac{\omega_j}{\omega_k}}\big)_\infty \ .
\end{align}

The {\bf Nekrasov function} is defined as
\be
N_{\lambda\mu}(x;\q,\t^{-1})\equiv \prod_{(i,j)\in\lambda}(1-x\q^{\lambda_i-j}\t^{\mu^\vee_j-i+1})\prod_{(i,j)\in\mu}(1-x\q^{-\mu_i+j-1}\t^{-\lambda^\vee_j+i})~,
\ee
where ${^\vee}$ denotes transposition of the Young diagrams.

  \section{Derivation of the $\S^3_{(1)}\, \cup\, \S^3_{(2)}$ matrix model}\label{app:computations}

Here we sketch how to derive the matrix model (\ref{matrixmodel}) following the argument given above (\ref{eq:1loopRes}). The exact equality between the residue of the $\S^5$ integrand at the selected poles (\ref{poles}) (with $n^{(3)}=0$) and the $\S^3_{(1)}\cup\S^3_{(2)}$ matrix model is established in the next section in the notation used in the main body. See also \cite{Pan:2016fbl} for another derivation.
\begin{figure}[t]
  \centering
  \includegraphics[width=0.3\textwidth]{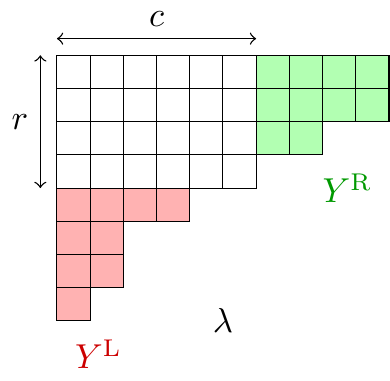}
  \caption{\label{youngdia}}
\end{figure}

We start by rewriting the instanton sum (\ref{eq:Z1}) using the manipulations considered in \cite{Nieri:2017ntx}. Shown in figure \ref{youngdia} is a \textit{large} Hook Young diagram $\lambda$ decomposed into an upper-left full rectangle with exactly $r$ rows and $c$ columns, an upper-right sub-diagram $Y^\text{R}$ with at most $r$ rows and a lower-left sub-diagram $Y^\text{L}$ with \textit{at most} $c$ rows. For such a diagram, we can write the corresponding summand in the instanton partition function (\ref{eq:Z1}) as
\begin{multline}
\frac{N_{\emptyset \lambda}(Q_1 \p^{1/2}; \q,\t)N_{\lambda \emptyset}(Q_2 \p^{1/2}; \q,\t)}{N_{\lambda\lambda}(1; \q,\t)}=\frac{1}{\mathcal{N}_{\emptyset\emptyset}}\prod_{i=1}^r\prod_{j=1}^{c}\frac{(1-\p^{1/2} Q_2 \q^{j-1}\t^{1-i})(1-\p^{1/2} Q_1 \t^{i}\q^{-j})}{(1-\t^{i}\q^{j-1}) (1-\t^{1-i}\q^{-j})}\times\\
\times\Delta_{\t}(z_{Y^\text{R}};\q)\Delta_{\q^{-1}}(z_{Y^\text{L}};\t^{-1})\prod_{i\geq 1}\frac{(\eta_\text{R}^{-1} \t^r \q^{-c}z_{Y^\text{R}_i}/x;\q)_\infty}{(\t \eta_\text{R} \t^{-r}\q^c x/z_{Y^\text{R}_i};\q)_\infty}\frac{(\eta_\text{L}^{-1} \t^r \q^{-c} z_{Y^\text{L}_i}/x;\t^{-1})_\infty}{(\q^{-1} \eta_\text{L} \t^{-r}\q^c x/z_{Y^\text{L}_i};\t^{-1})_\infty} \times\\
\times \prod_{i,j\geq 1}\frac{1}{(1-\p^{-1/2} z_{Y^\text{L}_j}/z_{Y^\text{R}_i})(1-\p^{-1/2} z_{Y^\text{R}_i}/z_{Y^\text{L}_j})}\times\\
\times\prod_{i\geq 1}\frac{(\t \eta_\text{R} \p^{1/2} Q_1  x/z_{Y^\text{R}_i};\q)_\infty}{(\eta^{-1}_\text{R}\p^{1/2} Q_2  z_{Y^\text{R}_i}/x;\q)_\infty}\frac{(\q^{-1} \eta_\text{L} \p^{1/2}Q_1 x/z_{Y^\text{L}_i};\t^{-1})_\infty}{(\eta^{-1}_\text{L} \p^{1/2}Q_2  z_{Y^\text{L}_i}/x;\t^{-1})_\infty}~,
\end{multline}
where $\eta_{\text{L},\text{R}}$ are free parameters such that $\eta_\text{L}/\eta_\text{R}=\sqrt{\q\t}$, we defined
\be
z_{Y^\text{L}_i}\equiv \eta_\text{L}\, x\, \t^{-r}\, \q^{i-1}\, \t^{-Y^\text{L}_i}~,\quad z_{Y^\text{R}_i}\equiv \eta_\text{R}\, x\, \q^c\, \t^{1-i}\, \q^{Y^\text{R}_i}~,
\ee
and $\mathcal{N}_{\emptyset\emptyset}$ denotes the whole factor beginning in the second line and evaluated for empty diagrams. The non-perturbative instanton partition function is obtained as the weighted sum over $\lambda$ with weight $(\mathfrak{p}^{-1/2}Q_0)^{|\lambda|}$, where $|\lambda|\equiv \sum_i \lambda_i$ implies the total number of boxes in $\lambda$. The sum can be further decomposed into a form respecting the hook Young diagram decomposition as shown in Figure \ref{youngdia}, namely $\sum_\lambda=\sum_{r,c\geq 0}\sum_{Y^\text{L},Y^\text{R}}$, such that $r-c=\mathfrak{n}$ is a fixed arbitrary integer expressing a linear relation between $r$ and $c$. Note that if we tune $\p^{1/2}Q_2=\q^{-n_1}\t^{n_2}~,$ the first factor in (\ref{eq:Z1}) vanishes, and therefore the instanton sum only receives non-vanishing contributions from diagrams $\lambda$ which do not contain the box $(n_2+1,n_1+1)$, i.e. Hook diagrams with $\lambda_{n_2+1} \le n_1$, $\lambda^\vee_{n_1+1}\leq n_2$: they include all large hook Young diagrams with an upper-left rectangle of the shape $r = n_2, c = n_1$, and infinitely many diagrams that we call \textit{small} hook diagrams. Let us focus on the large Hook diagrams. In this case we get the simplification
\begin{multline}\label{one:simplified}
\frac{N_{\emptyset \lambda}(Q_1 \p^{1/2}; \q,\t)N_{\lambda \emptyset}(Q_2 \p^{1/2}; \q,\t)}{N_{\lambda\lambda}(1; \q,\t)}=\prod_{i=1}^r\prod_{j=1}^{c}\frac{(1- \q^{-j}\t^{i})(1-\p^{1/2} Q_1 \t^{i}\q^{-j})}{(1-\t^{i}\q^{j-1}) (1-\t^{1-i}\q^{-j})}\times\\
\times\frac{\Delta_{\t}(z_{Y^\text{R}};\q)\Delta_{\q^{-1}}(z_{Y^\text{L}};\t^{-1})}{\mathcal{N}_{\emptyset\emptyset}}\prod_{i\geq 1}\frac{(\t \eta_\text{R} \p^{1/2} Q_1  x/z_{Y^\text{R}_i};\q)_\infty}{(\t \eta_\text{R} \t^{-r}\q^c x/z_{Y^\text{R}_i};\q)_\infty}\frac{(\q^{-1} \eta_\text{L} \p^{1/2}Q_1 x/z_{Y^\text{L}_i};\t^{-1})_\infty}{(\q^{-1} \eta_\text{L} \t^{-r}\q^c x/z_{Y^\text{L}_i};\t^{-1})_\infty} \times\\
\times \prod_{i,j\geq 1}\frac{1}{(1-\p^{-1/2} z_{Y^\text{L}_j}/z_{Y^\text{R}_i})(1-\p^{-1/2} z_{Y^\text{R}_i}/z_{Y^\text{L}_j})}~.
\end{multline}
Also, the residue of the perturbative factor in (\ref{eq:Z1}) at a pole $\p^{1/2}Q_2=\q^{-n_1}\t^{n_2}$ reads
\be\label{5dpertvortex}
\prod_{i=1,2}\frac{1}{(\p^{1/2}Q_i;\q,\t^{-1})_\infty} \to \frac{{\rm Res}_{z=1}(z;\q,\t^{-1})^{-1}_\infty}{(\p^{1/2}Q_1;\q,\t^{-1})_\infty}\prod_{i=1}^r\prod_{j=1}^c\frac{1}{1-\q^{-j}\t^i}\prod_{j=1}^c\frac{1}{(\q^{-j};\t^{-1})_\infty}\prod_{i=1}^r\frac{1}{(\t^{i};\q)_\infty}~.
\ee
Notice that the second factor will cancel against the first factor in the numerator of (\ref{one:simplified}). We can also set
\be
Q_1=\q^c \t^{-r}\p^{1/2}w/x~,
\ee
and redefine 
\be
z_{Y^\text{L}_i}\q^{-c}\t^r=\eta_\text{L}\,\q^{-1}\, x\, \q^{-c+i}\, \t^{-Y^\text{L}_i}\to z_{Y^\text{L}_i} ~,\quad z_{Y^\text{R}_i}\q^{-c}\t^r= \eta_\text{R}\,\t\, x\, \t^{r-i}\, \q^{Y^\text{R}_i}\to z_{Y^\text{R}_i}~,
\ee
so that 
\begin{multline}
\frac{N_{\emptyset \lambda}(Q_1 \p^{1/2}; \q,\t)N_{\lambda \emptyset}(Q_2 \p^{1/2}; \q,\t)}{N_{\lambda\lambda}(1; \q,\t)}=\prod_{i=1}^r\prod_{j=1}^{c}\frac{(1- \q^{-j}\t^{i})(1-\p  \t^{i-r}\q^{c-j}w/x)}{(1-\t^{i}\q^{j-1}) (1-\t^{1-i}\q^{-j})}\times\\
\times\frac{\Delta_{\t}(z_{Y^\text{R}};\q)\Delta_{\q^{-1}}(z_{Y^\text{L}};\t^{-1})}{\mathcal{N}_{\emptyset\emptyset}}\prod_{i\geq 1}\frac{(\t \eta_\text{R} \p w  /z_{Y^\text{R}_i};\q)_\infty}{(\t \eta_\text{R}  x/z_{Y^\text{R}_i};\q)_\infty}\frac{(\q^{-1} \eta_\text{L} \p w /z_{Y^\text{L}_i};\t^{-1})_\infty}{(\q^{-1} \eta_\text{L}  x/z_{Y^\text{L}_i};\t^{-1})_\infty} \times\\
\times \prod_{i,j\geq 1}\frac{1}{(1-\p^{-1/2} z_{Y^\text{L}_j}/z_{Y^\text{R}_i})(1-\p^{-1/2} z_{Y^\text{R}_i}/z_{Y^\text{L}_j})}~.
\end{multline}
For convenience, we can also set
\be
\t\eta_\text{R}\p w \equiv w_\text{R}~,\quad \q^{-1}\eta_\text{L}\p w\equiv w_\text{L}~,\quad \t\eta_\text{R} x\equiv x_\text{R}~,\quad \q^{-1}\eta_\text{L} x\equiv x_\text{L}~,  
\ee
so that 
\begin{multline}\label{quasivortex}
\frac{N_{\emptyset \lambda}(Q_1 \p^{1/2}; \q,\t)N_{\lambda \emptyset}(Q_2 \p^{1/2}; \q,\t)}{N_{\lambda\lambda}(1; \q,\t)}=\prod_{i=1}^r\prod_{j=1}^{c}\frac{(1- \q^{-j}\t^{i})(1-\p  \t^{i-r}\q^{c-j}w/x)}{(1-\t^{i}\q^{j-1}) (1-\t^{1-i}\q^{-j})}\times\\
\times\frac{\Delta_{\t}(z_{Y^\text{R}};\q)\Delta_{\q^{-1}}(z_{Y^\text{L}};\t^{-1})}{\mathcal{N}_{\emptyset\emptyset}}\prod_{i\geq 1}\frac{(w_\text{R}  /z_{Y^\text{R}_i};\q)_\infty}{( x_\text{R}/z_{Y^\text{R}_i};\q)_\infty}\frac{(w_\text{L} /z_{Y^\text{L}_i};\t^{-1})_\infty}{( x_\text{L}/z_{Y^\text{L}_i};\t^{-1})_\infty} \times\\
\times \prod_{i,j\geq 1}\frac{1}{(1-\p^{-1/2} z_{Y^\text{L}_j}/z_{Y^\text{R}_i})(1-\p^{-1/2} z_{Y^\text{R}_i}/z_{Y^\text{L}_j})}~.
\end{multline}
Notice that
\begin{align}
\frac{\Theta(\xi \p^{-1/2}Q_0/z_{Y^\text{R}_i};\q)\Theta(\xi ;\q)}{\Theta(\xi/ z_{Y^\text{R}_i};\q)\Theta(\xi \p^{-1/2}Q_0;\q)}&=\frac{\Theta(\xi \p^{-1/2}Q_0/z_{\emptyset^\text{R}_i};\q)\Theta(\xi ;\q)}{\Theta(\xi/ z_{\emptyset^\text{R}_i};\q)\Theta(\xi \p^{-1/2}Q_0;\q)}\, (\p^{-1/2}Q_0)^{|Y^\text{R}_i|}~,\\
\frac{\Theta(\xi \p^{-1/2}Q_0/z_{Y^\text{L}_i};\t^{-1})\Theta(\xi ;\t^{-1})}{\Theta(\xi/ z_{Y^\text{L}_i};\t^{-1})\Theta(\xi \p^{-1/2}Q_0;\t^{-1})}&=\frac{\Theta(\xi \p^{-1/2}Q_0/z_{\emptyset^\text{L}_i};\t^{-1})\Theta(\xi ;\t^{-1})}{\Theta(\xi/ z_{\emptyset^\text{L}_i};\t^{-1})\Theta(\xi \p^{-1/2}Q_0;\t^{-1})}\, (\p^{-1/2}Q_0)^{|Y^\text{L}_i|}~,
\end{align}
where $\xi$ is arbitrary. Since
\be
(\p^{-1/2}Q_0)^{|\lambda|}=(\p^{-1/2}Q_0)^{rc}(\p^{-1/2}Q_0)^{|Y^\text{L}|}(\p^{-1/2}Q_0)^{|Y^\text{R}|}~,
\ee
we can recognize the weighted sum over the left and right diagrams (second and third line of (\ref{quasivortex})) as the vortex part of the partition function 
\begin{align}\label{blockint}
\mathcal{B}_{\text{LR}}&\equiv\oint\prod_{i=1}^r\frac{\d z_{\text{R}i}}{2\pi\i z_{\text{R}i}}\prod_{j=1}^c\frac{\d z_{\text{L}j}}{2\pi\i z_{\text{L}j}}\, \Upsilon_\text{L}(z_\text{L})\Upsilon_{\rm int}(z_\text{L},z_\text{R})\Upsilon_\text{R}(z_\text{R})=\\
&={\rm Res}_{\substack{z_{\text{L}j}=z_{\emptyset^\text{L}_j}\\ z_{\text{R}i}=z_{\emptyset^\text{R}_i}}}\Upsilon_\text{L}(z_\text{L})\Upsilon_{\rm int}(z_\text{L},z_\text{R})\Upsilon_\text{R}(z_\text{R})\sum_{Y^\text{L},Y^\text{R}}\frac{\Upsilon_\text{L}(z_{Y^\text{L}})\Upsilon_{\rm int}(z_{Y^\text{L}},z_{Y^\text{R}})\Upsilon_\text{R}(z_{Y^\text{R}})}{\Upsilon_\text{L}(z_\emptyset^\text{L})\Upsilon_{\rm int}(z_\emptyset^\text{L},z_\emptyset^\text{R})\Upsilon_\text{R}(z_\emptyset^\text{R})}~,
\end{align}
where
\begin{align}
\Upsilon_\text{R}(z_\text{R})&\equiv\prod_{i=1}^r\frac{\Theta(\xi \p^{-1/2}Q_0/z_{\text{R}i};\q)\Theta(\xi ;\q)}{\Theta(\xi/ z_{\text{R}i};\q)\Theta(\xi \p^{-1/2}Q_0;\q)}\Delta_{\t}(z_\text{R};\q)\prod_{i=1}^r\frac{(w_\text{R}  /z_{\text{R}i};\q)_\infty}{( x_\text{R}/z_{\text{R}i};\q)_\infty}~,\\
\Upsilon_\text{L}(z_\text{L})&\equiv\prod_{j=1}^c\frac{\Theta(\xi \p^{-1/2}Q_0/z_{\text{L}j};\t^{-1})\Theta(\xi ;\t^{-1})}{\Theta(\xi/ z_{\text{L}j};\t^{-1})\Theta(\xi \p^{-1/2}Q_0;\t^{-1})}\Delta_{\q^{-1}}(z_\text{L};\t^{-1})\prod_{j=1}^c\frac{(w_\text{L} /z_{\text{L}j};\t^{-1})_\infty}{( x_\text{L}/z_{\text{L}j};\t^{-1})_\infty} ~,\\
\Upsilon_{\rm int}(z_\text{L},z_\text{R})&\equiv\prod_{i=1}^r\prod_{j=1}^c\frac{1}{(1-\p^{-1/2} z_{\text{L}j}/z_{\text{R}i})(1-\p^{-1/2} z_{\text{R}i}/z_{\text{L}j})}~,
\end{align}
and the contour is chosen to encircle the poles\footnote{We simply integrate the $z$'s one after the other, starting from $z_{R,i=r}$ around $x_\text{R}$ and $z_{L,j=c}$ around $x_\text{L}$.}
\be
z_{\text{L}j}=z_{Y^\text{L}_j}=x_\text{L}\, \q^{-c+j}\, \t^{-Y^\text{L}_j}~,\quad z_{\text{R}i}=z_{Y^\text{R}_i}= x_\text{R} \t^{r-i}\, \q^{Y^\text{R}_i}~.
\ee
This corresponds to the block integral \cite{Beem:2012mb} of the $\text{SQCDA-}\U(r)\cup\text{SQCDA-}\U(c)$ theory on $[\C_\q\times\S^1]\cup [\C_{\t^{-1}}\times\S^1]$, interacting through a pair of 1d chiral multiplets in the bi-fundamental of $\U(r)\times\U(c)$ at the common $\S^1$ intersection at the origin (plus superpotential terms). The 3d FI/vortex counting parameters $\zeta_\text{L}$, $\zeta_\text{R}$ are identified with 
\be
\p^{-1/2}Q_0=\q^{\zeta_\text{R}}=\t^{-\zeta_\text{L}}~.
\ee

Now let us think of $\C_\q\times\S^1$ and $\C_{\t^{-1}}\times\S^1$ as two halves of two squashed $\S^3$'s, namely
\be
\S^3_{(1)}\simeq [\C_\q\times\S^1] \#_S[ \C_{\tilde\q}\times\S^1]~,\quad \S^3_{(2)}\simeq [\C_{\t^{-1}}\times\S^1] \#_S[  \C_{\tilde\t^{-1}}\times\S^1]~,
\ee 
where $\tilde\q$ and $\tilde\t$ are related to $\q$ and $\t$ by the $S$ element in $\text{SL}(2,\mathbb{Z})$ performing the boundary homeomorphism \cite{Beem:2012mb}, and form the partition function on the intersecting space $\S^3_{(1)}\cup\S^3_{(2)}$. In order to do that, it is convenient to parametrize the variables as
\begin{align}
\q&\equiv\e^{2\pi\i\frac{\omega_1}{\omega_3}}~,\quad &\t^{-1}&\equiv\e^{2\pi\i\frac{\omega_2}{\omega_3}}~,\quad& \p&\equiv\e^{2\pi\i\frac{\rho}{\omega_3}}~, \nn\\
z_{\text{L}j}&\equiv\e^{\frac{2\pi\i}{\omega_3}Z_{\text{L}j}}~,\quad &
z_{\text{R}i}&\equiv\e^{\frac{2\pi\i}{\omega_3}Z_{\text{R}i}}~,\quad &
x_{\text{L},\text{R}}&\equiv\e^{\frac{2\pi\i}{\omega_3}X_{\text{L},\text{R}}}~,\quad &w_{\text{L},\text{R}}&\equiv\e^{\frac{2\pi\i}{\omega_3}W_{\text{L},\text{R}}}~,\nn\\
\p^{-1/2}Q_0&\equiv\e^{\frac{2\pi\i}{\omega_3}\zeta}~,\quad&  \xi&\equiv\e^{\frac{2\pi\i}{\omega_3}\Xi}~. 
\end{align}
Then we can multiply (\ref{blockint}) with another left block integral with $\omega_3\leftrightarrow \omega_2$ and another right block integral with $\omega_3\leftrightarrow\omega_1$. This will convert
\begin{align}
(\cdots;\q)_\infty&\to S_2(\cdots|\omega_{1},\omega_{3})\e^{-\frac{\i\pi}{2}B_{22}(\cdots|\omega_1,\omega_3)}~, &(\cdots;\t^{-1})_\infty&\to S_2(\cdots|\omega_{2},\omega_{3})\e^{-\frac{\i\pi}{2}B_{22}(\cdots|\omega_2,\omega_3)}~,\nn\\
\Theta(\cdots;\q)&\to \e^{-\i\pi B_{22}(\cdots|\omega_1,\omega_3)}~, &\Theta(\cdots;\t^{-1})&\to \e^{-\i\pi B_{22}(\cdots|\omega_2,\omega_3)}~.
\end{align}
Then the matrix model we are interested in becomes
\be
Z^{\S^3_{(1)}\cup\S^3_{(2)}}\equiv\int\d^c Z_\text{L}\d^r Z_\text{R}\, Z^{\S^3_{(2)}}_{\text{cl}}(Z_\text{L})Z^{\S^3_{(2)}}_{\text{1-loop}}(Z_\text{L})Z_{\text{int}}^{\S^1}(Z_\text{L},Z_\text{R})Z^{\S^3_{(1)}}_{\text{cl}}(Z_\text{R})Z^{\S^3_{(1)}}_{\text{1-loop}}(Z_\text{R})~,
\ee
where
\begin{align}
Z^{\S^3_{(1)}}_{\text{1-loop}}(Z_\text{R})&\equiv\prod_{1\leq i\neq j\leq r}\frac{S_2(Z_{\text{R}i}-Z_{\text{R}j}|\omega_1,\omega_3)}{S_2(-\omega_2+Z_{\text{R}i}-Z_{\text{R}j}|\omega_1,\omega_3)}\prod_{i=1}^r\frac{S_2(W_\text{R}-Z_{\text{R}i}|\omega_1,\omega_3)}{S_2(X_\text{R}-Z_{\text{R}i}|\omega_1,\omega_3)}~,\\
Z^{\S^3_{(2)}}_{\text{1-loop}}(Z_\text{L})&\equiv\prod_{1\leq i\neq j\leq c}\frac{S_2(Z_{\text{L}i}-Z_{\text{L}j}|\omega_2,\omega_3)}{S_2(-\omega_1+Z_{\text{L}i}-Z_{\text{L}j}|\omega_2,\omega_3)}\prod_{j=1}^c\frac{S_2(W_\text{L}-Z_{\text{L}j}|\omega_2,\omega_3)}{S_2(X_\text{L}-Z_{\text{L}j}|\omega_2,\omega_3)}~,\\
Z^{\S^3_{(1)}}_{\text{cl}}(Z_\text{R})&\equiv\e^{\frac{\i\pi\omega_2}{2\omega_1\omega_3}(r^2-1)(\omega_1+\omega_2+\omega_3)}\times\e^{-\frac{\i\pi}{2\omega_1\omega_3}r(W_\text{R}-X_\text{R})(W_\text{R}+X_\text{R}-\omega_1-\omega_3)}\times\nn\\
&~~~\times\prod_{i=1}^r \e^{\frac{\i\pi}{\omega_1\omega_3}(W_\text{R}-X_\text{R})Z_{\text{R}i}}\times \prod_{i=1}^r\e^{\frac{2\pi\i}{\omega_1\omega_3}\zeta Z_{\text{R}i}}~,\\
Z^{\S^3_{(2)}}_{\text{cl}}(Z_\text{L})&\equiv\e^{\frac{\i\pi\omega_1}{2\omega_2\omega_3}(c^2-1)(\omega_1+\omega_2+\omega_3)}\times\e^{-\frac{\i\pi}{2\omega_2\omega_3}c(W_\text{L}-X_\text{L})(W_\text{L}+X_\text{L}-\omega_2-\omega_3)}\times\nn\\
&~~~\times\prod_{j=1}^c \e^{\frac{\i\pi}{\omega_2\omega_3}(W_\text{L}-X_\text{L})Z_{\text{L}j}}\times \prod_{j=1}^c\e^{\frac{2\pi\i}{\omega_2\omega_3}\zeta Z_{\text{L}j}}~,\\
Z_{\text{int}}^{\S^1}(Z_\text{L},Z_\text{R})&\equiv\prod_{i=1}^r\prod_{j=1}^r\prod_\pm\frac{\e^{\frac{\i\pi}{\omega_3}\rho}}{4\sin\frac{\pi}{\omega_3}\Big[Z_{\text{R}i}-Z_{\text{L}j}\pm\frac{\rho}{2}\Big]}~.
\end{align}
Notice the renormalization of the FI by $(W_\text{L}-X_\text{L})/2=(W_\text{R}-X_\text{R})/2$ (we impose this equality), as usual when going from K-theoretic to field-theoretic notation. The vortex part of the above matrix model captures the Hook truncation of the $\S^5$ integrand at the poles specified in (\ref{poles}) with $n^{(3)}=0$. In order to obtain the exact equality between the matrix model and the residue of the $\S^5$ integrand at these poles, one needs to carefully study the extra factors in the first line of (\ref{quasivortex}), their combination with the 5d perturbative contributions (\ref{5dpertvortex}) as well as the residue of the matrix model at the trivial poles (perturbative part). Also, in order to fully specify the matrix model, one needs to choose an integration contour. The right choice turns out to be a Jeffrey-Kirwan prescription as studied in \cite{Pan:2016fbl}. Intuitively, the poles coming from the $\S^3$'s integrands will capture the contribution from large Hook diagrams (namely those constructed over a rectangle of size $r\times c$ and considered in this appendix), while the contribution from small Hook diagrams (namely those which do not contain the box $(r,c)$) are accounted by additional poles coming from the $\S^1$ piece.

\section{$\S^3$ and $\S_{(1)}^3 \cup \S_{(2)}^3$ partition functions}\label{app:S3partitionfunctions}
In this appendix, we establish the exact equality between the residue of the $\S^5$ integrand at the selected poles (\ref{poles}) (with $n^{(3)}=0$) and the $\S^3_{(1)}\cup\S^3_{(2)}$ matrix model (\ref{matrixmodel}) in the notation used in the main body. We start by recalling useful definitions of partition functions on a squashed spheres or their intersections. 

The squashed $\S^3$ partition function of a $\U(n)$ gauge theory coupled to $n_\text{f} = n_\text{af}$ fundamental and anti-fundamental chirals and one adjoint, which we will refer to as $\U(n)$-SQCDA, is given by
\begin{align}
  Z^{\S^3}_{\U(n)\text{-SQCDA}} \equiv & \ \int \frac{\d^n\sigma} {(2\pi \i)^n \, n!} \, \e^{-2\pi \i \zeta \sum_a\sigma_a} \prod_{a > b} 2 \sinh\pi b(\sigma_a - \sigma_b) 2 \sinh\pi b^{-1}(\sigma_a - \sigma_b) \nonumber\\
  & \ \times \prod_{i=1}^{n_\text{f}} \frac{
    \prod_{a=1}^n s_b(+ \i Q/2 + \sigma_a - \tilde m_i)
  }{
    \prod_{a=1}^n s_b( - \i Q/2 + \sigma_a - m_i)
  }
  \prod_{a,b=1}^n s_b\Big( \frac{\i Q}{2} - \sigma_a + \sigma_b + m_\text{adj} \Big) \ .
\end{align}
As usual, $b$ denotes the squashing parameter, $Q \equiv b + b^{-1}$, while $m_i$, $\tilde m_i$ and $m_{\rm adj}$ denote the complexified masses of fundamental, anti-fundamental and adjoint chiral multiplets
\begin{align}
  m \equiv m^\mathbb{R} - q\frac{\i Q }{2} \ , \qquad \tilde m \equiv \tilde m^\mathbb{R} + \tilde q\frac{\i Q}{2} \ ,\qquad m_{\rm adj} \equiv m_{\rm adj}^\mathbb{R} - q_{\rm adj}\frac{\i Q }{2}~,
\end{align}
and $\zeta$ is the FI parameter. Let us denote the integrand simply as $Z^{\S^3}_{\U(n)\text{-SQCDA}}(\sigma)$. Then the partition function of a pair of $\U(n^{(\alpha)})$-SQCDA on $\S^3_{(1)} \cup \S^3_{(2)}$, interacting through a pair of 1d bi-fundamental chiral multiplets at the intersection $\S^1=\S^3_{(1)} \cap \S^3_{(2)}$, is given by
\begin{multline}\label{mmatrixmmodel}
Z^{\S^3_{(1)} \cup \S^3_{(2)}}_{\U(n^{(1)})\text{-SQCDA} \, \cup\,  \U(n^{(2)})\text{-SQCDA}}\equiv\\
\equiv \int \prod_{\alpha=1}^2\prod_{a=1}^{n^{(\alpha)}}\frac{\d \sigma_a^{(\alpha)}}{2\pi \i \, n^{(\alpha)}!}\,  Z^{\S^3_{(1)}}_{\U(n^{(1)}), n_\text{f},n_\text{af}}(\sigma^{(1)})Z^{\S^1}_\text{1d chiral}(\sigma^{(1)}, \sigma^{(2)})Z^{\S^3_{(2)}}_{\U(n^{(2)}), n_\text{f},n_\text{af}}(\sigma^{(2)}) ~,
\end{multline}
where the contribution from the 1d chiral multiplets is captured by
\begin{align}
  Z^{\S^1}_\text{1d chiral}(\sigma^{(1)}, \sigma^{(2)}) = \prod_\pm \prod_{a=1}^{n^{(1)}}\prod_{b=1}^{n^{(2)}} \frac{1}{
    2\i \sinh \pi \big(
      b_{(1)} \sigma_a^{(1)} - b_{(2)} \sigma_a^{(2)} \pm \frac{\i}{2}(b_{(1)}^2 + b_{(2)}^2)
    \big)
  } \ . 
\end{align}

In general, the parameters in the two SQCDA are independent, however, when they are the worldvolume theories of intersecting codimension 2 defects in a bulk 5d $\mathcal{N} = 1$ theory, the masses are likely to be related due to 5d/3d superpotentials, which is indeed the case throughout our paper. For example, we have mass relations
\begin{align}
  b_{(1)}m^{(1)}_i - b_{(2)}m^{(2)}_i = \frac{\i }{2}(b_{(2)}^2 - b_{(1)}^2) \ .
\end{align}

The matrix model (\ref{mmatrixmmodel}) should be understood as a contour integral with a Jeffrey-Kirwan residue prescription. Take $n^{(1)} = 1, n^{(2)} = 1$ as an example. There are two sets of poles, the first of which is given by
\begin{align}
  \sigma^{(1)} = m^{(1)} - \i \mathfrak{m}^{(1)} b_{(1)} - \i \mathfrak{n}^{(1)} b_{(1)}^{-1} \ , \qquad \sigma^{(2)} = m^{(2)} - \i \mathfrak{m}^{(2)} b_{(2)} - \i \mathfrak{n}^{(2)} b_{(2)}^{-1}  \ ,
\end{align}
for all $\mathfrak{m}^{(\alpha)}, \mathfrak{n}^{(\alpha)} \ge 0$, while the second
\begin{align}
  \sigma^{(1)} = m^{(1)} - \i (-1) b_{(1)} - \i \mathfrak{n}^{(1)} b_{(1)}^{-1}, \qquad \sigma^{(2)} = m^{(2)} - \i \mathfrak{n}^{(2)} b_{(2)}^{-1} \ ,
\end{align}
for all $\mathfrak{n}^{(\alpha)} \ge 0$. Clearly, the second set come from the poles of $Z^{\S^1}_\text{1d chiral}$, since this set of poles satisfy
\begin{align}
  \sinh \pi \Big(b_{(1)}\sigma^{(1)} - b_{(2)}\sigma^{(2)} - \frac{\i }{2}(b_{(2)}^2 + b_{(1)}^2) \Big) = 0 \ ,
\end{align}
thanks to the mass relations mentioned above. With these definitions, the equality (\ref{matrixmodel}) and the master identities (\ref{master-equality}) can be explicitly verified (e.g. by using Mathematica).

\section{The refined topological vertex and DIM algebra}\label{sec:topvertex}
\subsubsection*{The refined topological vertex}
\begin{figure}[t]
  \centering
  \includegraphics[width=0.4\textwidth]{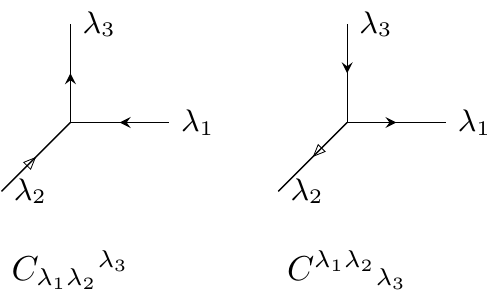}
  \caption{\label{vertices}Refined topological vertices.}
\end{figure}
The topological vertex formalism \cite{Aganagic:2003db} and its refinement \cite{Iqbal:2007ii,Awata:2008ed} are powerful tools to study 5d instanton partition functions and their properties. In this note we will mainly follow the conventions of \cite{Awata:2008ed}, which we now review.

The relevant vertices\footnote{There are two more vertices with different directions of the white arrows. However, we choose to build the web diagrams with just the two in Figure \ref{vertices}.} are graphically represented in Figure \ref{vertices}. Note that at each vertex there are two black and one white arrows (the preferred/instanton direction), each labeled by a Young diagram. The three arrows are ordered in a clockwise manner, keeping the white arrow in the middle. For example, in the two diagrams in the Figure \ref{vertices}, the white arrows are labeled with $\lambda_2$, and is also the second index of the vertex. Lowered/raised indices of the vertex correspond to incoming/outgoing arrows. These graphical vertices represent the following contributions to the full amplitute,
\begin{align}\label{Cvertex}
  & C_{\lambda_1 \lambda_2}{^{\lambda_3}} = P_{\lambda_2}(\t^\rho; \q,\t)\sum_\lambda \p^{\frac{|\lambda| - |\lambda_3|}{2}}f_{\lambda_3}(\q,\t)^{-1}\iota P_{\lambda_1^\vee/\lambda^\vee} (- t ^{\lambda^\vee} \q^\rho; \t,\q) P_{\lambda_3/\lambda}(\q^\lambda t^\rho; \q,\t) \ ,\\
  & C^{\lambda_1 \lambda_2}{_{\lambda_3}} = P_{\lambda_2^\vee}(- \q^\rho; \t,\q)\sum_\lambda
    \p^{\frac{|\lambda_3| - |\lambda|}{2}}
    f_{\lambda_3}(\q,\t)
    \iota P_{\lambda_1/\lambda} (\q^{\lambda} t^\rho; \q,\t) P_{\lambda_3^\vee/\lambda^\vee}( - \t^{\lambda^\vee} \q^\rho; \q,\t) \ .
\end{align}
The $P_{\lambda/\mu}(x;\q,\t)$ is the skew Macdonald function of the sequence of variables $x = (x_1, x_2, \ldots)$ with Young diagrams $\lambda = (\lambda_1, \lambda_2, \ldots)$ and $\mu = (\mu_1, \mu_2, \ldots)$ as parameters, while $|\lambda|\equiv\sum_i\lambda_i$ denotes the total number of boxes in the diagram $\lambda$ and $\iota$ is the involution $\iota(p_n)=-p_n$ acting on the power sums $p_n\equiv \sum_i x_i^n$. The other parameters $\q \equiv \e^{2\pi \i \epsilon_1}, \t \equiv \e^{- 2\pi \i \epsilon_2} $ and $\p \equiv \q\t^{-1}$ are complex numbers.

The vertices can be joined together to form web diagrams corresponding to CY or $(p,q)$-webs engineering 5d supersymmetric gauge theories. In doing so, each internal line is further associated to a complex parameter $Q^{|\lambda|}$ and a framing factor $f_{\lambda}(\q,\t)^{n}$ (for us $n=0$), and the corresponding Young diagrams are summed over.  

\subsubsection*{DIM intertwiners}
\begin{figure}[t]
  \centering
  \includegraphics[width=1\textwidth]{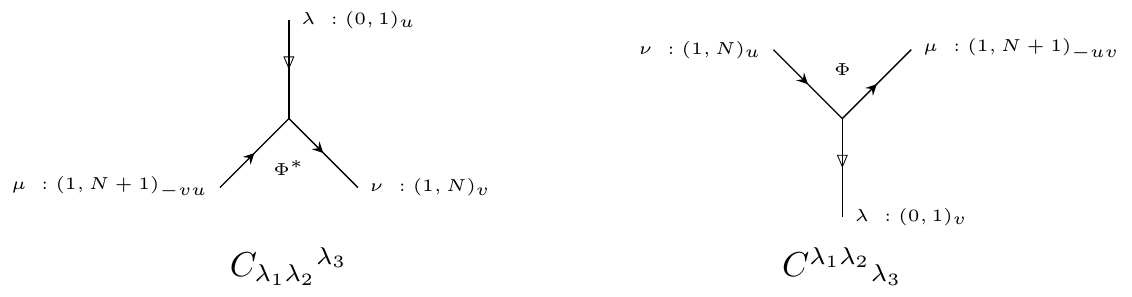}
  \caption{\label{DIMintert} DIM intertwining operators.}
\end{figure}

The topological vertex can be interpreted as matrix elements of DIM intertwining operators in the Macdonald basis \cite{Awata:2011ce}, namely
\begin{align}
C^{\mu\lambda}_{~~~\nu}(\q,\t)&=Q_{N_{(u,v)}}^{|\lambda|}(t^{-1/2}v)^{|\nu|-|\mu|}\frac{f_\lambda^{N}(\q,\t)f_\nu(\q,\t)}{\braket{ P_\lambda}{P_\lambda}}
\bra{\iota P_\mu}\Phi_\lambda\left[{\tiny {\arraycolsep=-2pt\begin{array}{ccc}&(1,N+1)_{-uv}&\\(0,1)_v&&(1,N)_u\end{array}}}\right]\ket{\iota Q_\nu}\nn\\
C_{\mu\lambda}^{~~~\nu}(\q,\t)&=Q_{N_{(v,u)}}^{-|\lambda|}(t^{-1/2}u)^{|\mu|-|\nu|}\frac{1}{f_\lambda^{N}(\q,\t)f_\nu(\q,\t)}
\bra{\iota P_\nu}\Phi^*_\lambda\left[{\tiny {\arraycolsep=-2pt\begin{array}{ccc}(1,N)_{v}&&(0,1)_u\\&(1,N+1)_{-uv}&\end{array}}}\right]\ket{\iota Q_\mu}~,
\end{align}
where we defined $Q_{N_{(x,y)}}\equiv -\q(-y)^N/\t^{1/2}x$. The state $\ket{\iota P_\mu}$ and its dual $\bra{\iota Q_\mu}$ give a Fock basis, and the labels $(n,k)_x$ are DIM representations specified by the integer values of the two central charges and the complex spectral parameter. In particular, $(0,1)_x$ is called vertical, while $(1,N)_x$ is called horizontal. They are isomorphic and related by the so-called spectral duality \cite{Mironov:2012uh,Mironov:2013xva,Awata:2016mxc}, a manifestation of the $\textrm{SL}(2,\mathbb{Z})$ group of automorphism of the DIM algebra. In the web diagram, the choice of preferred/white direction correspond to the choice of vertical representation, to which $\Phi$ or $\Phi^*$ are attached. See figure \ref{DIMintert} for an illustration.

\begin{figure}[t]
  \centering
  \includegraphics[width=1\textwidth]{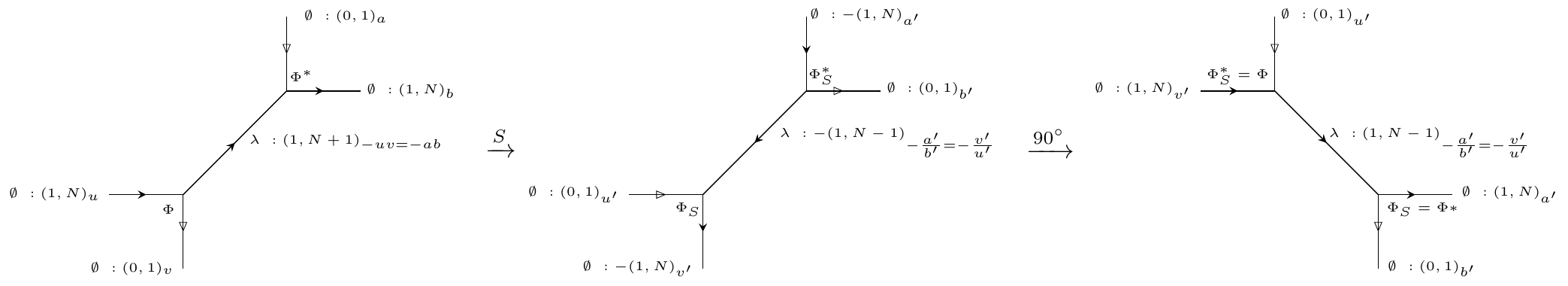}
  \caption{\label{90deg} Action of the $S$ element in $\textrm{SL}(2,\mathbb{Z})$ (for $N=0$).}
\end{figure}

\begin{figure}[t]
  \centering
  \includegraphics[width=0.6\textwidth]{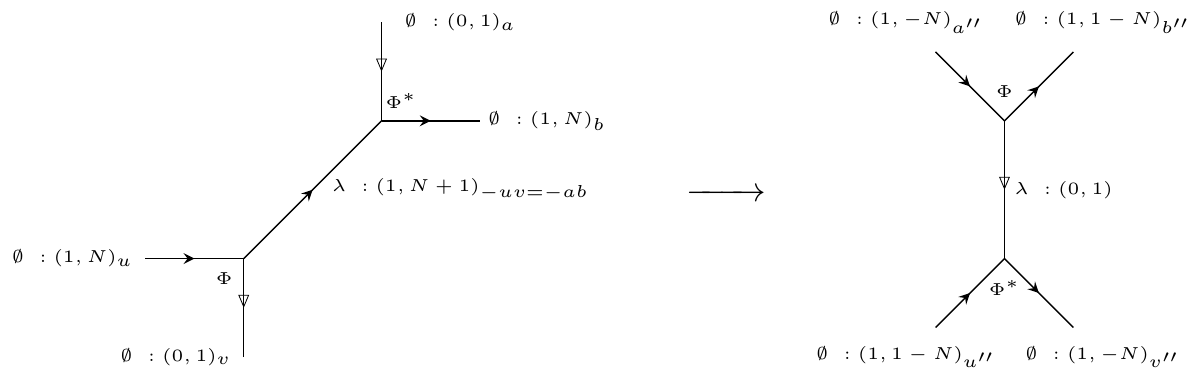}
  \caption{\label{45deg} The third triality frame for the resolved conifold.}
\end{figure}

As the basic example, let us consider the resolved conifold amplitude with preferred direction or $(0,1)$ representation along the vertical direction
\be
\bra{\emptyset}\Phi^*_\emptyset\left[{\tiny {\arraycolsep=-2pt\begin{array}{ccc}(1,N)_{b}&&(0,1)_a\\&(1,N+1)_{-ab}&\end{array}}}\right]\Phi_\emptyset\left[{\tiny {\arraycolsep=-2pt\begin{array}{ccc}&(1,N+1)_{-uv}&\\(0,1)_v&&(1,N)_u\end{array}}}\right]\ket{\emptyset}=\sum_{\lambda}(v/a)^{|\lambda|}C_{\lambda\emptyset}^{~~~\emptyset}(\q,\t) C^{\lambda\emptyset}_{~~~\emptyset}(\q,\t)~,
\ee
where $uv=ab$. Alternatively, we could have put the preferred direction or $(0,1)$ representation along the horizontal direction
\be
\bra{\emptyset}\Phi^*_\emptyset\left[{\tiny {\arraycolsep=-2pt\begin{array}{ccc}(1,N-1)_{-v'/u'}&&(0,1)_{u'}\\&(1,N)_{v'}&\end{array}}}\right]\Phi_\emptyset\left[{\tiny {\arraycolsep=-2pt\begin{array}{ccc}&(1,N)_{a'}&\\(0,1)_{b'}&&(1,N-1)_{-a'/b'}\end{array}}}\right]\ket{\emptyset}=\sum_{\lambda}(b'/u')^{|\lambda|}C_{\lambda\emptyset}^{~~~\emptyset}(\q,\t) C^{\lambda\emptyset}_{~~~\emptyset}(\q,\t)~,
\ee
where $a'/b=v'/u'$. The two results should agree because of slicing invariance of the topological vertex, and they do provided $v/a=b'/u'\equiv Q_0$, which is the ratio of the outgoing/incoming spectral parameters associated to the $(0,1)$ representations. From the DIM perspective, this should descend from the $\textrm{SL}(2,\mathbb{Z})$ automorphism of the algebra, see Figure \ref{90deg} for an illustration. 
A more complicated choice is to assign the preferred direction or $(0,1)$ representation to the diagonal direction. Now the composition of the intertwiners acts on the tensor product of two Fock spaces, and the corresponding amplitude is
\begin{multline}
\bra{\emptyset}\otimes\bra{\emptyset}\sum_\lambda\frac{1}{\braket{P_\lambda}{P_\lambda}}\Phi_\lambda\left[{\tiny {\arraycolsep=-2pt\begin{array}{ccc}&(1,1-M)_{b''}&\\(0,1)_{-b''/a''}&&(1,-M)_{a''}\end{array}}}\right]\otimes \Phi^*_\lambda\left[{\tiny {\arraycolsep=-2pt\begin{array}{ccc}(1,-M)_{v''}&&(0,1)_{-u''/v''}\\&(1,1-M)_{u''}&\end{array}}}\right]\ket{\emptyset}\otimes\ket{\emptyset}=\\
=\sum_\lambda(a''/v'')^{|\lambda|}C^{\emptyset\lambda}_{~~~\emptyset}(\q,\t)C_{\emptyset\lambda}^{~~~\emptyset}(\q,\t)~,
\end{multline}
where $b''/a''=u''/v''$. This corresponds to the Nekrasov partition function of the 5d pure $\textrm{U}(1)$ SYM  theory with instanton counting parameter $a''/v''$. This expansion coincides with the previous ones provided we identify $a''/v''=Q_0$. See figure \ref{45deg} for an illustration.

For the next level of complication, we can consider the geometries considered in the main text. As we discussed, there is a frame corresponding to a $\textrm{U}(1)$ theory with two flavors (Figure \ref{three-diagrams} left), a frame corresponding to four free hypers (Figure \ref{three-diagrams} center) and a frame corresponding to a $\textrm{U}(1)\times \textrm{U}(1)$ theory with one bi-fundamental hyper (Figure \ref{three-diagrams} right). It is now easy to recognize the various topological amplitudes as (vacuum) matrix elements of intertwining operators between various representations, and the fact that they should agree is expected from the $\textrm{SL}(2,\mathbb{Z})$ automorphism of DIM. In particular, we can identify (we neglect the unnecessary labels in order to avoid clutteing)
\begin{align}
Z_1&=\bra{\emptyset}\otimes\bra{\emptyset}\sum_{\lambda}\frac{\left(\Phi^*_\emptyset\otimes\mathds{1}\right)\left(\Phi_\lambda\otimes\Phi^*_\lambda\right)\left(\mathds{1}\otimes\Phi_\emptyset\right)}{\braket{P_\lambda}{P_\lambda}}\ket{\emptyset}\otimes\ket{\emptyset}~,\\
Z_2&=\bra{\emptyset}\Phi^*_\emptyset\Phi_\emptyset\Phi^*_\emptyset\Phi_\emptyset\ket{\emptyset}~,\\
Z_3&=\bra{\emptyset}\otimes\bra{\emptyset}\otimes\bra{\emptyset}\sum_{\lambda_1,\lambda_2}\frac{\left(\mathds{1}\otimes\Phi_{\lambda_1}\otimes\Phi^*_{\lambda_1}\right)\left(\Phi_{\lambda_2}\otimes\Phi^*_{\lambda_2}\otimes\mathds{1}\right)}{\braket{P_{\lambda_1}}{P_{\lambda_1}}\braket{P_{\lambda_2}}{P_{\lambda_2}}}\ket{\emptyset}\otimes\ket{\emptyset}\otimes\ket{\emptyset}~.
\end{align}
Of course, we need suitable identifications between parameters. Anyhow, from the form of the matrix elements it is immediate that $Z_1$ should correspond to a $\textrm{U}(1)$ theory, $Z_2$ to a free theory and $Z_3$ to a $\textrm{U}(1)\times \textrm{U}(1)$ theory. Also, since the W$_{\q,\t^{-1}}(A_1)$ or $\q$-Virasoro algebra can be represented on the tensor product of two horizontal DIM representations, while W$_{\q,\t^{-1}}(A_2)$ can be represented on the tensor product of three horizontal DIM representations, the resulting 5d $\mathcal{N}=1$ quiver gauge theories match with Kimura-Pestun construction of quiver W$_{\q,\t^{-1}}$ algebras \cite{Kimura:2015rgi}. In their construction, the basic object is the $\mb{Z}$ operator, which is an \textit{infinite} product of the W$_{\q,\t^{-1}}$ screening charges. From the DIM perspective, we can identify
\be
\mb{Z}[A_1]=\sum_{\lambda}\frac{\Phi_\lambda\otimes\Phi^*_\lambda}{\braket{P_\lambda}{P_\lambda}}~,\quad \mb{Z}[A_2]=\sum_{\lambda_1,\lambda_2}\frac{\left(\mathds{1}\otimes\Phi_{\lambda_1}\otimes\Phi^*_{\lambda_1}\right)\left(\Phi_{\lambda_2}\otimes\Phi^*_{\lambda_2}\otimes\mathds{1}\right)}{\braket{P_{\lambda_1}}{P_{\lambda_1}}\braket{P_{\lambda_2}}{P_{\lambda_2}}}~.
\ee
On the other hand, it is known that Kimura-Pestun construction as an analogous for 3d $\mathcal{N}=2$ quiver gauge theories, which involves a \textit{finite} number of   W$_{\q,\t^{-1}}$ screening charges \cite{Aganagic:2013tta,Aganagic:2014kja,Aganagic:2014oia,Nedelin:2016gwu}. An efficient control on the transformation relations between the DIM operators in different duality frames and at specific points in the parameter space (corresponding to complete Higgsing of the 5d theories) would imply an elegant description of some 3d dualities. The peculiar example of the self-mirror $T[\U(N)]$ theory \cite{Gaiotto:2008ak} has been recently considered in \cite{Zenkevich:2017ylb} from the W$_{\q,\t^{-1}}$ perspective.

\end{appendix}

\bibliographystyle{utphys}
\providecommand{\href}[2]{#2}\begingroup\raggedright\endgroup

\end{document}